\documentclass[aps,prb,twocolumn,floatfix]{revtex4-1}
\usepackage{graphicx,txfonts,bm}
\usepackage[colorlinks,citecolor=magenta,linkcolor=blue]{hyperref}

\begin{document}


\title{Nature of the 5$f$ electronic structure of plutonium}


\author{Li Huang}
\email{lihuang.dmft@gmail.com}
\affiliation{Science and Technology on Surface Physics and Chemistry Laboratory, P.O. Box 9-35, Jiangyou 621908, China}

\author{Haiyan Lu}
\affiliation{Science and Technology on Surface Physics and Chemistry Laboratory, P.O. Box 9-35, Jiangyou 621908, China}

\date{\today}


\begin{abstract}
Plutonium (Pu), in which the 5$f$ valence electrons always wander the boundary between localized and itinerant states, exhibits quite complex crystal structures and unprecedentedly anomalous properties with respect to temperature and alloying. Understanding its chemical and physical properties, especially its 5$f$ electronic structure is one of the central and unsolved topics in condensed matter theory. In the present work, the electronic structures of the six allotropes of Pu (including its $\alpha$, $\beta$, $\gamma$, $\delta$, $\delta'$, and $\epsilon$ phases) at ambient pressure are studied comprehensively by means of the density functional theory in combination with the single-site dynamical mean-field theory. The band structures, total and partial density of states, valence state histograms, 5$f$ orbital occupancies, X-ray branching ratios, and self-energy functions are carefully studied. It is suggested that the $\alpha$, $\beta$, and $\gamma$ phases of Pu are typical Racah metals in which the atomic multiple effect dominates near the Fermi level. The calculated results reveal that not only the $\delta$ phase, but also all the six allotropes are archetypal mixed-valence metals with remarkable atomic eigenstate fluctuation. In consequence of that, the 5$f$ occupancy $n_{5f}$ is around 5.1 $\sim$ 5.4, which varies with respect to the atomic volume and electronic correlation strength of Pu. The 5$f$ electronic correlation in Pu is moderately orbital-dependent. Moreover, the 5$f$ electrons in the $\delta'$ phase are the most correlated and localized.
\end{abstract}


\maketitle


\section{Introduction\label{sec:intro}}

Plutonium is a radioactive element with atomic number 94 and the chemical symbol Pu. As is well-known, Pu, the sixth member of the actinide series, is considered to be one of the most mysterious, complex, and exotic elements in the periodic table~\cite{handbook}. It is an element at odds with itself. Some peoples indeed claim that metallic Pu is a physicist's dream but an engineer's nightmare (because it defies conventional metallurgical wisdom)~\cite{HECKER2004429,LAReview}. It has attracted a lot of interests and studies since its discovery at 1940. To date, there are still tons of questions and puzzles concerning its unusual properties that need to be answered and solved~\cite{LAReview,entropy:2019,nh:2019,joyce:2019,rt:2019,Migliori11158}.   

\begin{figure*}[ht]
\centering
\includegraphics[width=\textwidth]{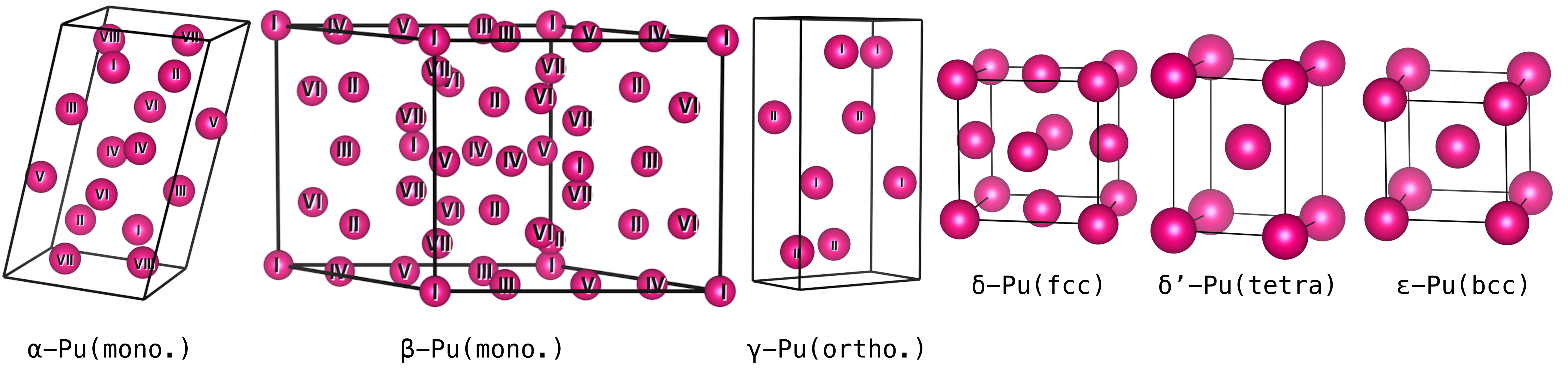}
\caption{(Color online). Schematic lattice structures of Pu's six allotropes~\cite{handbook} studied in the present work. The crystalline structures are enclosed in the parentheses. Here mono., ortho., tetra., fcc, and bcc are abbreviations of monoclinic, orthorhombic, tetragonal, face-centered cubic, and body-centered cubic, respectively. The non-equivalent sites in the low-symmetry $\alpha$-Pu, $\beta$-Pu, and $\gamma$-Pu phases are depicted using Roman numerals (I $\sim$ VIII)~\cite{Zachariasen:a03908,Zachariasen:a02472,Zachariasen:a01450}. \label{fig:tstruct}}
\end{figure*}

Plutonium's $V-T$ phase diagram is extremely complicated, and (at ambient pressure) comprises six allotropes which have different crystal structures (see Fig.~\ref{fig:tstruct}) and manifest distinct lattice properties~\cite{Hecker2004,LAReview,HECKER2004429}. These allotropes can be roughly classified into two categories according to their crystal structures and symmetries: (1) Low-symmetry $\alpha$, $\beta$, and $\gamma$ phases~\cite{handbook,HECKER2004429,LAReview}. Under ambient temperature and pressure, the $\alpha$ phase is favorable. It crystallizes in a monoclinic structure with 16 Pu atoms within the unit cell~\cite{Zachariasen:a03908}. These Pu atoms can be grouped into eight non-equivalent types (Pu$_{\alpha,\text{I}}$ $\sim$ Pu$_{\alpha,\text{VIII}}$). The crystal structure of $\beta$-Pu, which is stable at higher temperature, is also monoclinic but with even more atoms (34 Pu atoms) within the unit cell~\cite{Zachariasen:a02472}. They are grouped into seven non-equivalent types (Pu$_{\beta,\text{I}}$ $\sim$ Pu$_{\beta,\text{VII}}$). The crystal structure of the orthorhombic $\gamma$ phase is less complex than those of $\alpha$- and $\beta$-Pu, but its unit cell still contains two non-equivalent Pu atoms (Pu$_{\gamma,\text{I}}$ $\sim$ Pu$_{\gamma,\text{II}}$)~\cite{Zachariasen:a01450}. (2) High-symmetry $\delta$, $\delta'$, and $\epsilon$ phases~\cite{handbook,HECKER2004429,LAReview}. The $\delta$, $\delta'$, and $\epsilon$ phases crystallize in the cubic and tetragonal structures, respectively. There is only one Pu atom in their unit cells. They are usually stable under elevated temperature. Finally, we note that Pu could form the seventh phase (i.e. $\zeta$-Pu) under high temperature and a limited pressure range~\cite{handbook,HECKER2004429}. 

As mentioned above, plutonium shows incredible sensitivity to temperature and demonstrates unusual lattice properties~\cite{handbook,LAReview,albers:2001}. When heated in the $\alpha$ phase, it expands at a rate almost five times the rate in iron. On the contrary, it contracts while being heated in the $\delta$ phase~\cite{HECKER2004429}. Because its liquid phase is denser than the previous solid phase, it contracts while melting at $T > 913$~K. In addition, its liquid state exhibits the greatest viscosity of any element and a very high surface tension emerges~\cite{PhysRevE.83.026404}. Pu shows even more atypical behaviors once it is cooled down below room temperature. It is a poor electrical conductor with very high electrical resistivity at room temperature. However, its resistivity increases gradually as the temperature is lowered to 100~K. It is also a bad thermal conductor and its specific heat is ten times larger than normal value at temperatures close to 0~K~\cite{PhysRevLett.91.205901,PhysRevLett.92.146401}. The magnetic susceptibility at low temperature is unusually high and almost retains constant, being a signature of magnetism. But in spite of that, so far, none of the long-range ordering states (including magnetic and superconducting) have been observed in Pu even at the lowest temperatures~\cite{PhysRevB.72.054416}.   

Perhaps $\delta$-Pu is the most useful and familiar phase due to its broad applications in the military and energy industries. Yet it could be the least understood theoretically phase due to its confusing and fascinating characteristics~\cite{handbook,LAReview,albers:2001}. It crystallizes in a cubic close-packed structure, but its density is the lowest. From brittle $\alpha$ phase to ductile $\delta$ phase, the volumetric change is up to record-high 25\%~\cite{HECKER2004429,smith:198383}. It usually stabilizes under high temperature, but doping it with a few percent trivalent metal impurities (such as Ga and Al) makes it metastable at room temperature~\cite{handbook,LAReview,HECKER2004429}. The $\delta$ phase has negative thermal expansion coefficient, unlike the vast majority of metallic materials~\cite{PhysRevB.81.214402,zpyin:2014,invar:2006}. The Pauli-like magnetic susceptibility, electrical resistivity, and Sommerfeld coefficient of the specific heat of $\delta$-Pu are an order of magnitude larger than those of the other simple metals~\cite{PhysRevB.5.4564,PhysRevB.72.054416,PhysRevLett.91.205901}. Additionally, the lattice dynamics of $\delta$-Pu is notoriously peculiar~\cite{dai:2003,wong:2003,PhysRevLett.92.146401,PhysRevB.79.052301,PhysRevB.58.15433}. In the calculated and experimental phonon dispersion curves of $\delta$-Pu, the $T[111]$ mode exhibits a pronounced bending along the transverse branch, which can be related to its lattice instability and the $\delta-\epsilon$ phase transition~\cite{HECKER2004429}. The longitudinal and transverse acoustic phonon branches along [001] direction are nearly degenerate at small wave vector $\vec{q}$, which leads to approximately equal elastic constants $C_{11}$ and $C_{44}$. In consequence, the $\delta$ phase shows astonishing shear anisotropy, much larger than those in any other simple face-centered cubic metals known~\cite{dai:2003,wong:2003}.

It is generally believed that electronic structure determines nonnuclear properties. Hence, in order to explain why plutonium metal behaves so strangely, we have to understand its electronic structure at first\cite{albers:2001,Hecker2004,RevModPhys.81.235}. Plutonium belongs to the actinides. The actinides successively fill the 5$f$ shell, much like the rare earths fill the 4$f$ shell~\cite{LAReview}. There is no doubt that the Janus-faced 5$f$ electronic structures (being itinerant or localized) are responsible for a plethora of interesting physical behaviors of the actinides~\cite{albers:2001,LAReview,RevModPhys.81.235,PhysRevB.76.115116}. In the early (light) actinides (from Ac to Np), their 5$f$ electrons behave much more like the 4$d$ or 5$d$ electrons of the transition metals, instead of the 4$f$ electrons of the lanthanides. They incline to be itinerant and contribute to chemical bonding~\cite{nature:1995}. There are at least three consequences for the itinerant 5$f$ electrons. Firstly, the 5$f$ electrons occupy the conduction band, which leads to an increase in chemical bonding force and a decrease in atomic volume~\cite{smith:198383}. Secondly, there are no local moments. The third, the 5$f$ electrons in the light actinides usually form very narrow and nearly flat energy bands, which manifest themselves by very high density of states near the Fermi level. Actually, the bonding properties of the light actinides are dominated by the specific properties of these flat bands. For example, low-symmetry structures are favored in the ground states of light actinides. This is because lattice distortions can split these narrow 5$f$ bands and thereby lower the total energy~\cite{LAReview}. In the late (heavy) actinides (from Am to No), the scenario looks a bit different. Their 5$f$ electrons start to be localized at each lattice sites and become chemically inert, behaving like the 4$f$ electrons of the rare earths~\cite{smith:198383}. The localized 5$f$ electrons usually give rise to nontrivial local magnetic moments. In addition, atomic volumes of the late actinides only shrink slightly with increasing atomic number, because the 5$f$ electrons in the remaining of the series come to be more and more localized. Pu sits halfway across the row of actinides. It happens to separate the early and late actinides. The electronic structure of Pu may be unique in the periodic table~\cite{LAReview,Hecker2004}. The reasons are two-folds. On one hand, its 5$f$ valence electrons live at the brink between localized and itinerant configurations. On the other hand, the degree of 5$f$ electron localization strongly depends on crystal structure and external conditions, such as temperature, stress, and chemical doping (alloying). Right at plutonium, there appears to be a 5$f$ itinerant-(partially) localized transition or crossover between the monoclinic $\alpha$ phase and face-centered cubic $\delta$ phase~\cite{Joyce2006920,LAReview}. Furthermore, it is concluded that the 5$f$ electronic structures of the six allotropes of Pu are completely diverse~\cite{RevModPhys.81.235} and the six phases of Pu are virtually different metals~\cite{PhysRevB.84.064105,Hecker2004}.  

The fundamental nature of 5$f$ electrons is at the research frontier of condensed matter physics. The 5$f$ electronic structure is critically essential to the structural and mechanical properties of plutonium, particularly to its phase transition and phase stability~\cite{PhysRevB.84.064105,HECKER2004429,Hecker2004}. Unfortunately, except for the $\delta$ phase, our knowledge about the electronic structures of the other phases is quite insufficient. Generally speaking, the electronic structure of Pu remains actually unexplained. An unified picture for the evolution of electronic structures of all of six phases of Pu with respect to temperature and crystal symmetry is highly desired. Keeping these deficiencies in mind, we try to study the electronic structures of the six allotropes of Pu by using a state-of-the-art first-principles many-body approach, i.e., the combination of density functional theory and dynamical mean-field theory (dubbed as DFT + DMFT)~\cite{RevModPhys.78.865}. In the present work, we elaborate the tendency of 5$f$ electrons from itinerant to partial localization that occurs in the different phases of Pu. We also identify some atypical features, such as atomic multiplets, valence state fluctuations, and orbital-dependent correlations, which are totally unexpected. Our findings suggest that the complexity of 5$f$ electronic structures is far away from being fully understood. 

The rest of this paper is organized as follows. In Sec.~\ref{sec:review}, previously theoretical results concerning with Pu's 5$f$ electronic structure are briefly reviewed. Two different computational strategies (with or without 5$f$ electronic correlation effect in the calculations) are summarized and discussed. In Sec.~\ref{sec:method}, we firstly introduce the spirit and advantages of our first-principles many-body computational framework (the DFT + DMFT method). And then we supplement the computational parameters and details. Sec.~\ref{sec:results} is the major part of this paper. In this section we present the theoretical electronic structures (including band structures, density of states, atomic eigenstate histograms, X-ray absorption branching ratios, 5$f$ occupancies, and 5$f$ self-energy functions) for all allotropes of Pu under ambient pressure. In Sec.~\ref{sec:compare}, the calculated results are compared with the available experimental and theoretical data. In Sec.~\ref{sec:discuss}, three important issues [namely (i) the evolution of 5$f$ electron localization in the six allotropes, (ii) the possible influence of truncation approximation and negative sign problem during the DFT + DMFT calculations for electronic structures, (iii) site-dependent 5$f$ electronic structures for inequivalent Pu atoms] are discussed at first in detail. The similarities in the electronic structures of Ce and Pu are then summarized and emphasized. Finally, Section~\ref{sec:summary} serves as a brief conclusion and outlook.

\section{Brief review of previous results\label{sec:review}}

On one hand, the density functional theory (DFT) and its extensions are considered as work horses for most condensed-matter calculations. On the other hand, Pu has been regarded as one of the hardest tentative systems for \emph{ab initio} electronic structures calculations~\cite{soder_review,RevModPhys.81.235}. Therefore, extensive first-principles methods (mainly DFT and beyond DFT methods) have been invented and then employed to make progress toward this ``holy grails" in the past decades. Since these works are too numerous to cite, only the most relevant and important advances are reviewed here.       

The key issue in the theoretical calculations concerning Pu's electronic structure is about how to treat the strong correlation effect among its 5$f$ electrons. So, according to this criterion, we can put the available theoretical works into two sets approximately: ignoring the 5$f$ electronic correlations (traditional DFT methods) or considering them in the calculations explicitly (beyond DFT or DFT + $X$ methods)~\cite{soder_review}.   

\emph{Without 5$f$ electronic correlations: DFT calculations.} Although relativistic DFT calculations within local density approximation were already performed to study the electronic structure of Pu in the 1970s-1990s~\cite{achandbook,PhysRevB.43.14414}, the DFT method with generalized gradient approximation was first applied to plutonium by Per S\"{o}derlind \emph{et al.} in 1994~\cite{PhysRevB.50.7291}. He and his collaborators insist that spin-polarized DFT calculations, with orbital polarization and spin-orbit coupling, are capable of capturing Pu's phase diagram and yielding the nontrivial crystal structures of low-temperature phases. They have made remarkable achievements in understanding the crystal structures, magnetism, chemical bonding, elastic properties, lattice dynamics, phase transition and phase stability of Pu~\cite{PhysRevB.66.205109,PhysRevB.68.241101,PhysRevB.70.144103,PhysRevB.72.205122,PhysRevB.79.104110,PhysRevB.81.224110,PhysRevLett.92.185702,PhysRevB.94.115148,per:2015sr,per:2017sr,soder_review}. For example, they successfully reproduced the highly complex crystal structure and 13 independent elastic constants of $\alpha$-Pu, the anomalously soft $C'$ as well as a large anisotropy ratio ($C_{44}/C'$) of $\delta$-Pu~\cite{dai:2003,wong:2003}. They also proposed a simple model which is universally valid for all Pu's phases. This model establishes a relationship between atomic volume (density), crystal structure (symmetry), and magnetic moments. They further developed a new mechanism to explain why Ga can stabilize face-centered cubic $\delta$-Pu under room temperature and ambient pressure~\cite{PhysRevLett.96.206402,PhysRevLett.92.185702,soder_review}.

\emph{With 5$f$ electronic correlations: DFT + X calculations.} In order to take the 5$f$ electronic correlation (which is a typical many-body effect) into consideration, the single-particle picture of the DFT approach is not valid any more. Clearly, we need more powerful guns. If the on-site Coulomb interaction among strongly correlated electrons, parameterized by using the Coulomb repulsive interaction parameter $U$ and Hund's exchange interaction parameter $J_{\text{H}}$, is treated in a static and mean-field level, it is the so-called DFT + $U$ approach~\cite{jpcm:1997}. Since it can capture the correlated nature of the open 5$f$ shell, it has been successfully applied to a large number of actinide compounds. Note that it is usually in favor of a magnetic solution~\cite{Bouchet_2000,PhysRevB.72.024458,PhysRevLett.84.3670}. A. B. Shick \emph{et al.} have employed the DFT + $U$ approach to study the ground state properties of $\delta$-Pu. Surprisingly, they obtained a completely non-magnetic ground state for $\delta$-Pu as well as for Pu-Am alloys when reasonable values of $U$ (3 $\sim$ 4 eV) are adopted in the calculations~\cite{Shick_2005}. Boris Dorado \emph{et al.} combined the DFT + $U$ method and temperature-dependent effective potential (TDEP) method to study the lattice vibrational properties of the high-temperature $\delta$ and $\epsilon$ phases of plutonium~\cite{PhysRevB.95.104303}. They found that the $\epsilon$ phase can only be stabilized when the temperature and electronic correlation effects are simultaneously accounted for. Besides the DFT + $U$ approach, the DFT + DMFT method is another powerful approach to tackle the 5$f$ electron-electron interaction~\cite{RevModPhys.78.865}. We will introduce its basic principles in next section. Here, we would like to emphasize that DFT + DMFT may be the most commonly used method to study all aspects of Pu and the other actinides. For example, the valence fluctuation behaviors in $\delta$-Pu and Pu-Am alloys~\cite{shim:2007,Janoscheke:2015,PhysRevLett.101.126403}, subtle electronic structures of $\alpha$-Pu, $\beta$-Pu, $\delta$-Pu, and even the Pu-Ga alloy in its $\delta$ phase~\cite{savrasov:2001,PhysRevB.75.235107,Shick_2007,PhysRevLett.101.056403,zhu:2013,PhysRevB.76.245118,PhysRevB.91.165126,PhysRevB.97.039903,PhysRevB.99.125113}, electronic specific heat of $\alpha$-Pu and $\delta$-Pu~\cite{PhysRevB.75.235107}, high-temperature phonon spectra of $\delta$-Pu and $\epsilon$-Pu~\cite{dai:2003,Amadon_2018}, etc., were quite reasonably described within the framework of the DFT + DMFT approach. The Gutzwiller approximation in combination with the density functional theory (dubbed as DFT + $G$) also enables us to study complex 4$f$ and 5$f$ systems beyond the single-particle approximation~\cite{PhysRevB.85.035133,PhysRevB.79.075114}. Nicola Lanat\'{a} \emph{et al.} have employed this approach to study the zero temperature phase diagram and electronic structure of Pu, finding good agreement with the experiments~\cite{PhysRevX.5.011008}. They further argued that, it is the competition between the Peierls effect and the Madelung interaction, leading to the differentiation between the equilibrium densities of Pu's six allotropes. The dependence of the 5$f$ electron correlations on the lattice structure has a negligible effect.

\emph{With 5$f$ electronic correlations: $\text{GW}$ and $\text{QSGW}$ calculations.} We note that none of the above DFT + $X$ approaches is actually parameter-free. They at least require the input of on-site Coulomb interactions. On top of that, they also suffer an uncertainty about the double counting term problem~\cite{jpcm:1997,PhysRevLett.115.196403}. The diagrammatically based approaches provide alternative route to overcome these problems. Therefore, there is a significant interest in developing and using the GW approximation and its extensions, such as quasiparticle self-consistent GW (QSGW) method~\cite{PhysRevB.89.035104,PhysRevB.76.165106}. Andrey Kutepov \emph{et al.} have implemented a self-consistent fully relativistic GW method and applied it to study the $\delta$ phase of Pu~\cite{PhysRevB.85.155129}. They found that the GW approximation renormalized to spin-orbit split $5f_{5/2}$ and $5f_{7/2}$ states. Compared to the DFT, the $5f-6d$ hybridization in Pu is greatly enhanced by GW. A. Svane \emph{et al.} have applied the QSGW approach to the different phases of elemental Pu. They found a ``universal'' scaling relationship, specifically, the local density approximation band width is proportional to the $f$-electron band width reduction, which can be used to quantify the electronic correlation strength of Pu~\cite{PhysRevB.87.045109,chantis:2009}. 

In summary, we review briefly recent advances in realistic calculations of the complex electronic structure of Pu. Here, we survey a series of major methods (including DFT, DFT + $U$, DFT + $G$, DFT + DMFT, GW and QSGW). They describe Pu's 5$f$ electronic structure with increasing level of complexity at increasing computational cost, and yield a lot of exciting insights in the field of plutonium science. Further developments are underway to improve the accuracy, speed, and predictive power of these methods. 


\section{Method\label{sec:method}}

\begin{table*}[htbp]
\caption{Key parameters used in the present DFT + DMFT calculations. In this table, the settings for $k$-points ($k$-mesh), radius of Muffin-tin sphere ($R_{\text{MT}}$), size of basis set ($R_{\text{MT}}K_{\text{MAX}}$), exchange-correlation functional (XC), double counting term (DC), Coulomb repulsive interaction ($U$), Hund's exchange interaction ($J_{\text{H}}$), spin-orbital coupling constant ($\lambda_{\text{SO}}$), system temperature ($T$), and number of Monte Carlo sweeps ($N_{\text{sweeps}}$) per DMFT iteration (one-shot CT-HYB quantum impurity solver calculation) are shown. Here PBE means the Perdew-Burke-Ernzerhof functional~\cite{PhysRevLett.77.3865} and FLL means the fully localized limit scheme~\cite{jpcm:1997}. See main text for more explanations. \label{tab:param}}
\begin{ruledtabular}
\begin{tabular}{ccccccccccc}
cases & 
$k$-mesh &
$R_{\text{MT}}$ &
$R_{\text{MT}}K_{\text{MAX}}$ &
XC &
DC &
$U$ &
$J_{\text{H}}$ &
$\lambda_{\text{SO}}$ &
$T$ &
$N_{\text{sweeps}}$ \\
\hline
$\alpha$-Pu   & $11 \times 14 \times 06$ & 2.41 & 8.0 & PBE & FLL & 5.0 eV & 0.6 eV & 0.22   & 290\ K & 1.0 $\times 10^8$ \\
$\beta$-Pu    & $09 \times 10 \times 09$ & 2.44 & 8.0 & PBE & FLL & 5.0 eV & 0.6 eV & 0.22   & 464\ K & 1.0 $\times 10^8$ \\
$\gamma$-Pu   & $10 \times 10 \times 10$ & 2.50 & 8.0 & PBE & FLL & 5.0 eV & 0.6 eV & 0.22   & 527\ K & 1.0 $\times 10^8$ \\
$\delta$-Pu   & $15 \times 15 \times 15$ & 2.50 & 9.0 & PBE & FLL & 5.0 eV & 0.6 eV & 0.22   & 645\ K & 4.0 $\times 10^8$ \\
$\delta'$-Pu  & $17 \times 17 \times 17$ & 2.50 & 9.0 & PBE & FLL & 5.0 eV & 0.6 eV & 0.22   & 750\ K & 4.0 $\times 10^8$ \\
$\epsilon$-Pu & $15 \times 15 \times 15$ & 2.50 & 9.0 & PBE & FLL & 5.0 eV & 0.6 eV & 0.22   & 829\ K & 4.0 $\times 10^8$ \\
\end{tabular}
\end{ruledtabular}
\end{table*}

To account for the 5$f$ electron correlation effect, sophisticated quantum many-body algorithms are preferred. One of the most successful algorithms may be the dynamical mean-field theory (DMFT), which is based on a mapping of lattice models onto quantum impurity models subject to a self-consistency equation~\cite{RevModPhys.68.13}. This mapping is exact for lattice models in the limit of infinite spatial dimensions. Notice that in DMFT the spatial fluctuations are frozen, but the local quantum fluctuations are taken into accounts. The electronic self-energy is therefore momentum-independent. DMFT is very successful and has widespread applications in studying strongly correlated models. To become material specific, DMFT must be merged with DFT and then a new electronic structure tool (DFT + DMFT) is developed. In the framework of DFT + DMFT approach, DFT is responsible for the non-interaction orbitals and providing a band picture, while DMFT provides a non-perturbative treatment for the strongly correlated problems~\cite{RevModPhys.78.865}. The DFT + DMFT method has achieved great success in numerous correlated materials. It is really appropriate for exploring electronic structures, especially band structures and spectroscopic quantities, of strongly correlated materials. In the present work, we decide to choose the DFT + DMFT approach to study the electronic structures of the six allotropes of Pu thoroughly.

The DFT calculations were done by using the \texttt{WIEN2K} code~\cite{wien2k}, which implements a full-potential linear augmented plane-wave (FP-LAPW) formalism. We used the experimental lattice structures~\cite{handbook,HECKER2004429}, and conducted only paramagnetic calculations. The spin-orbit coupling effect was explicitly considered in the calculations. The most important computational parameters are summarized and listed in Table~\ref{tab:param}.   

We employed the \texttt{EDMFTF} software package, which was developed by K. Haule \emph{et al.}~\cite{PhysRevB.81.195107}, to do the DMFT calculations. The merit of this code is that it preserves stationarity of the DFT+DMFT functional, and is able to obtain high precision total energy and force~\cite{PhysRevLett.115.256402}. The hybridization expansion version of continuous-time quantum Monte Carlo quantum impurity solver (dubbed as CT-HYB) was used to solve the resulting multi-orbital (seven-band) impurity models~\cite{PhysRevLett.97.076405,RevModPhys.83.349}. In order to minimize the computational resources required, we considered some approximations and tricks. First, we assumed that the Pu atoms in $\alpha$-, $\beta$-, and $\gamma$-Pu are completely equivalent. In order words, we ignored the non-equivalent Pu atoms. Second, we utilized some good quantum numbers (such as total occupancy $N$ and total angular momentum $J$) to reduce the maximum matrix size of local impurity Hamiltonian. Third, we retained those atomic eigenstates with $N \in [3,7]$ only~\cite{PhysRevB.75.155113}. The other atomic eigenstates were discarded. Finally, we adopted the Lazy trace evaluation trick to accelerate the Monte Carlo sampling procedure further~\cite{PhysRevB.90.075149}. We carried out charge fully self-consistent DFT + DMFT calculations. About 60 $\sim$ 80 DFT + DMFT iterations are enough to obtain good convergence on charge, chemical potential, and total energy. Once the calculations are converged, we used the maximum entropy method to accomplish the analytical continuation~\cite{jarrell}, and tried to calculate the physical observables. The technical details are illustrated in Refs.~[\onlinecite{PhysRevB.81.195107}] and [\onlinecite{PhysRevB.99.045122}]. 


\section{Results\label{sec:results}}

\subsection{Stripe-like band structures}

\begin{figure*}[ht]
\centering
\includegraphics[width=\textwidth]{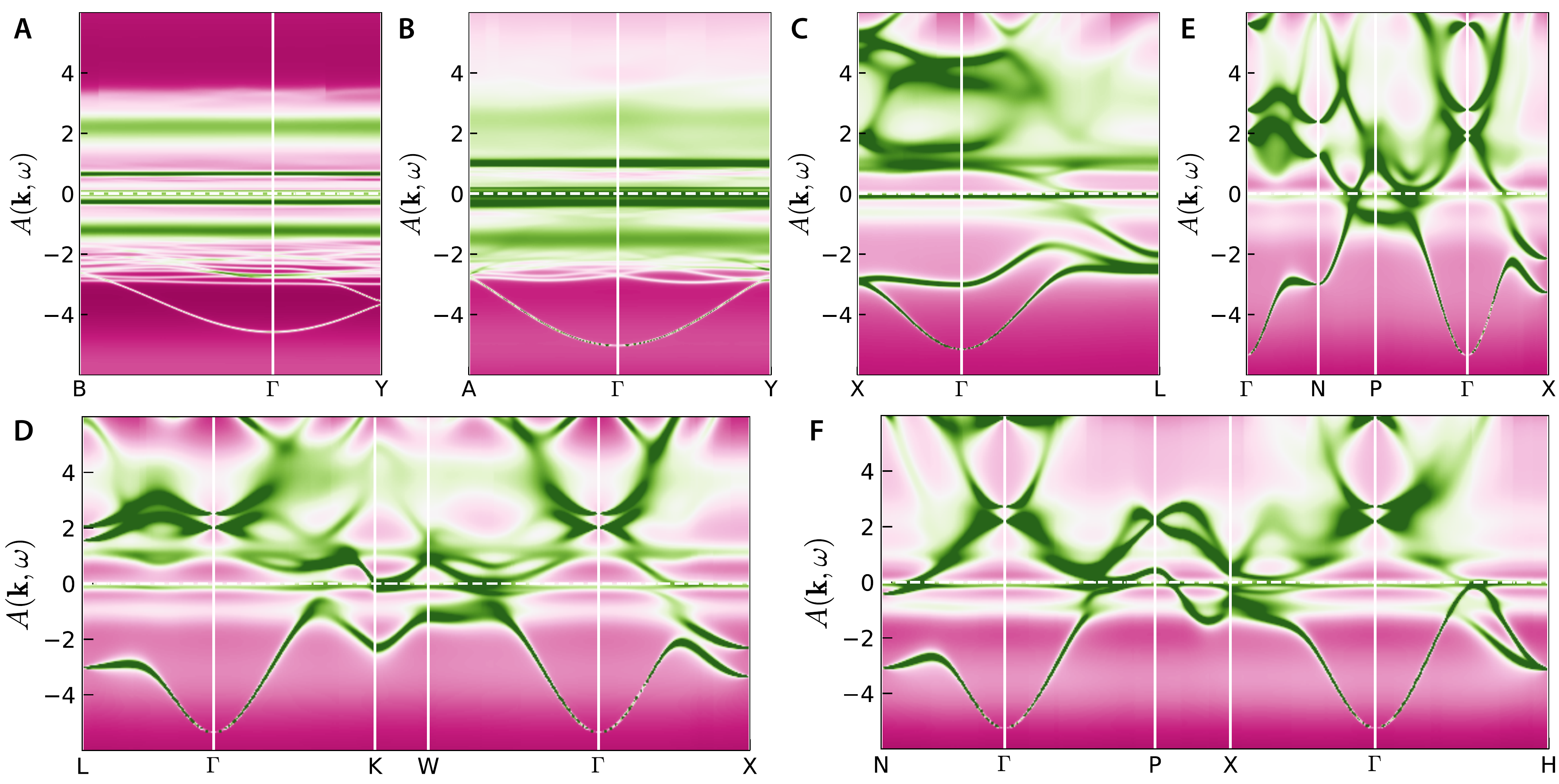}
\caption{(Color online). Momentum-resolved spectral functions $A(\mathbf{k},\omega)$ of Pu obtained by the DFT + DMFT method. (a) $\alpha$-Pu. The coordinates for the high-symmetry points are $B$[0.5, 0.0, 0.0], $Y$[0.0, 0.0, 0.5]. (b) $\beta$-Pu. The coordinates for the high-symmetry points are $A$[0.5, 0.0, 0.0], $Y$[0.0, 0.5, 0.0]. (c) $\gamma$-Pu. The coordinates for the high-symmetry points are $X$[0.0, 0.0, 1.0], $L$[0.5, -0.5, 0.5]. (d) $\delta$-Pu. (e) $\delta$'-Pu. (f) $\epsilon$-Pu. The horizontal dashed lines denote the Fermi levels. \label{fig:takw}}
\end{figure*}

The momentum-resolved spectral functions $A(\mathbf{k},\omega)$ is an ideal theoretical tool to observe directly the distribution of electrons in the reciprocal space of solids, which provides deep insights into the valence electrons of correlated electron materials. The corresponding experimental technique is the angle-resolved photoemission spectroscopy (ARPES). Here, we illustrate the calculated results, $A(\mathbf{k},\omega)$ along some selected high-symmetry lines in the irreducible Brillouin zone, for plutonium in Fig.~\ref{fig:takw}. The following characteristics are noticeable: (i) As for $\alpha$- and $\beta$-Pu, the most striking features are the parallel and intensive stripe-like patterns in the spectra. For instance, these stripes locate approximately at -1.2 eV, -0.2 eV, 0.6 eV, and 2.0 eV for $\alpha$-Pu (-1.5 eV, -0.2 eV, and 1.0 eV for $\beta$-Pu). These stripes probably resemble the 5$f$ atomic multiplets. Since there are too many non-equivalent Pu atoms in the unit cell (see Fig.~\ref{fig:tstruct}), it seems the spectra are quite blurry and somewhat overcrowded. Besides these stripes, it is difficult to find out any other special features and identify the hybridization gaps. (ii) For $\gamma$-Pu, there are also apparently stripe-like patterns in the spectrum. Their positions are close to those in the $\alpha$ and $\beta$ phases, but with smaller intensity. In addition, since there are only two non-equivalent Pu atoms in the unit cell (see Fig.~\ref{fig:tstruct}), more features (band dispersions) can be identified in the band structures. We observe prominent $c-f$ hybridizations along the $L-\Gamma$ line in the Brillouin zone. (iii) For $\delta$-, $\delta'$-, and $\epsilon$-Pu, they are all high-temperature phases with high-symmetry crystal structures (only one Pu atom in the unit cell, see Fig.~\ref{fig:tstruct}). It is worth saying that their spectra are amazingly similar. All of them show quite clear band dispersions from -4.0 eV to -0.5 eV and from 2.0 eV to 4.0 eV. In the vicinity of the Fermi level and around $\pm 1.0$\ eV, there are dim and almost flat band structures, which are very likely associated with the partially localized 5$f$ bands.

From the momentum-resolved spectral functions of Pu, one could confirm that the 5$f$ electrons in the high-temperature phases ($\delta$-, $\delta'$-, and $\epsilon$-Pu) are partially localized~\cite{RevModPhys.81.235}, and might speculate roughly that the electronically localized degrees of freedom are quite different for various phases~\cite{PhysRevB.84.064105}. However, the most prominent thing is that the 5$f$ electrons in $\alpha$-, $\beta$-, and $\gamma$-Pu are not well described with the itinerant electron picture, which is in sharp contrast to our expectation. Recently, A. Svane \emph{et al.} have proposed a new variable $\mathcal{C}$ to quantify the electronic correlation strengths of all six allotropic phases of Pu~\cite{PhysRevB.87.045109,chantis:2009}. They calculated $\mathcal{C}$ via the following equations:
\begin{equation}
\mathcal{C} = 1 - \omega_{\text{rel}},
\end{equation}
and
\begin{equation}
\omega_{\text{rel}} = \frac{W^{5f}_{\text{QSGW}}}{W^{5f}_{\text{LDA}}},
\end{equation}
where $\omega_{\text{rel}}$ is the relative band width reduction in the QSGW approximation~\cite{PhysRevLett.96.226402,PhysRevB.74.245125,PhysRevB.76.165106} compared to DFT in the local density approximation (LDA)~\cite{PhysRevB.23.5048}, $W^{5f}_{\text{QSGW}}$ and $W^{5f}_{\text{LDA}}$ are the $5f$ band widths as obtained from QSGW and DFT(LDA) calculations, respectively. They found that $\mathcal{C}_{\alpha} < \mathcal{C}_{\beta} < \mathcal{C}_{\gamma} < \mathcal{C}_{\epsilon} < \mathcal{C}_{\delta'} \approx \mathcal{C}_{\delta}$. In other words, the $\delta$ (or $\delta'$) phase has the largest correlation strength and the most localized 5$f$ electrons. Apparently, our calculations aren't in complete accordance with their results. At least, the 5$f$ electronic correlation strengths in the $\alpha$, $\beta$, and $\gamma$ phases are suspicious. Lastly, to our knowledge, it is the first time to obtain the $A(\mathbf{k},\omega)$ for Pu via \emph{ab initio} many-body approach. Consequently, it would be very helpful to examine them via high-resolution photoemission spectroscopy and quantum oscillation experiment in the future~\cite{rt:2019,joyce:2019}.

\subsection{Atomic multiplets}

\begin{figure*}[ht]
\centering
\includegraphics[width=\textwidth]{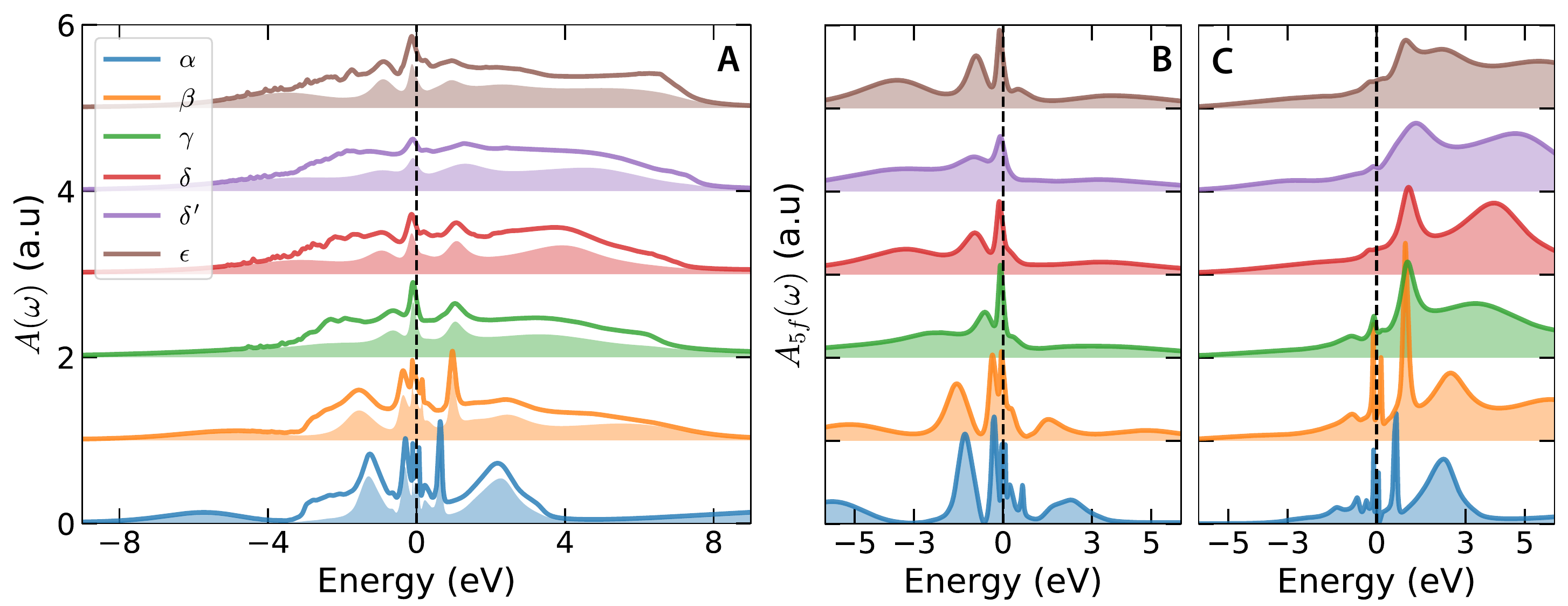}
\caption{(Color online). Total and 5$f$ partial density of states of Pu by DFT + DMFT calculations. (a) Total density of states $A(\omega)$ (in solid lines) and $5f$ partial density of states $A_{5f}(\omega)$ (in colored shadow regions). (b) and (c) Orbital-resolved (or $j$-resolved) $5f$ partial density of states $A_{5f_{5/2}}(\omega)$ and $A_{5f_{7/2}}(\omega)$. The Fermi levels $E_{\text{F}}$ are represented by vertical dashed lines. Note that the spectral data in this figure have been rescaled and normalized for a better visualization. \label{fig:tdos}}
\end{figure*}

Now let us turn to the total and 5$f$ partial density of states of plutonium, $A(\omega)$ and $A_{5f}(\omega)$, which can be regarded as the results of momentum integral of $A(\mathbf{k},\omega)$:
\begin{equation}
A(\omega) = \int_{\Omega} d\mathbf{k} A(\mathbf{k},\omega).
\end{equation}
The calculated results are illustrated in Fig.~\ref{fig:tdos}. For the $\alpha$ and $\beta$ phases, there exist several sharp and intensive peaks near the Fermi level, which are related with the stripe-like features as already observed in the momentum-resolved spectral functions (see Fig.~\ref{fig:takw}). For the $\gamma$ phase, there are still multiple peaks on the verge of the Fermi level, but the total band width is greatly reduced with comparison to the $\alpha$ and $\beta$ phases. For the $\delta$, $\delta'$, and $\epsilon$ phases, their 5$f$ partial density of states are quite similar. All of them show fat and short quasi-particle resonance peaks (accompanying with broad and smooth Hubbard bands at high energy regime), instead of atomic multiplets at the Fermi level. 

It is generally believed that Pu's 5$f$ electrons sit at the edge of an itinerant-localized transition, where small changes or perturbations will result in a transition to itinerancy or localization~\cite{RevModPhys.81.235,albers:2001,Hecker2004}. C. A. Marianetti \emph{et al.} have employed the DFT + DMFT method to calculate the volume dependence of magnetic susceptibility and temperature dependence of the valence band photoemission spectra of $\delta$-Pu~\cite{PhysRevLett.101.056403}. They found that expanding the volume would drive the 5$f$ electrons in $\delta$-Pu to crossover from coherent to incoherent state at increasingly lower temperatures. At high temperatures, the spectra are diffuse with small weights at the Fermi level. As the temperature is decreased, a quasi-particle peak continually builds and finally saturates~\cite{PhysRevLett.101.056403}. In order to analyze the evolution of 5$f$ electronic structures of the six allotropes of Pu, we try to evaluate the 5$f$ integrated spectral weights near the Fermi level:
\begin{equation}
\label{eq:w5f}
I_{5f} = \int^{+\Delta}_{-\Delta} A_{5f}(\omega) d\omega,
\end{equation}
where $\Delta = 0.2$~\cite{limits}. Before the calculations, $A_{5f}(\omega)$ has been normalized to satisfy the sum-rule. We find that the calculated values of $I_{5f}$ satisfy the following relations:
\begin{equation}
\label{eq:i5f}
I_{5f}(\alpha) \approx I_{5f}(\beta) \approx I_{5f}(\gamma) > I_{5f}(\delta) \approx I_{5f}(\delta') \approx I_{5f}(\epsilon).
\end{equation} 
This trend is compatible with the change in lattice volume of Pu as a function of temperature~\cite{HECKER2004429,PhysRevX.5.011008}. Actually, $\alpha$-Pu has the smallest lattice volume per Pu atom (19.5 $\sim$ 20.4 \AA$^3$). For $\delta$-Pu, its lattice volume per Pu atom is the largest (25.0 $\sim$ 25.5 \AA$^3$)~\cite{handbook}. If we plot the $I_{5f}$ against the atomic volumes of the $\alpha$, $\beta$, $\gamma$, $\delta$, $\delta'$, and $\epsilon$ phases, the resulting curve is approximately a Heaviside step function. Our calculations manifest that $I_{5f}$ may be a good measurement for the electronic coherence in various phases of Pu, and is equivalent to $\mathcal{C}$ proposed by A. Svane \emph{et al.}~\cite{PhysRevB.87.045109,chantis:2009} in some extent. According to our calculated results, we believe that the 5$f$ electrons are coherent in the low-temperature $\alpha$, $\beta$, and $\gamma$ phases, and tend to be incoherent in the high-temperature $\delta$, $\delta'$, and $\epsilon$ phases. 

The distinguishing feature for the density of states of the $\alpha$, $\beta$, and $\gamma$ phases is the coexistence of atomic-like quasi-particle resonance peaks near the Fermi level and itinerant-like Hubbard bands at high energy regime. These quasi-particle peaks mainly originate from the many-body transitions between the $5f^6$ and 5$f^5$ atomic multiplet configurations, while the Hubbard bands are related to the $5f^4-5f^5$ transitions~\cite{PhysRevB.81.035105}. Let us further inspect the main quasi-particle peak at the Fermi level. It has a small quasi-particle weight $Z$ (and narrow band width), which implies that it is strongly renormalized as compared to the DFT (LDA) density of states and the electron effective mass is quite large (see Table~\ref{tab:weight}). Additionally, it is very sensitive to the variation of temperature~\cite{PhysRevLett.101.056403}. With decreasing temperature it sharpens, considerably reducing the density of states at the Fermi level and leading to the formation of a ``pseudogap''. However, upon increasing temperature it merges with the multiplet peaks at the Fermi edge gradually. As a consequence, for the $\delta$, $\delta'$, and $\epsilon$ phases, we only observe a single quasi-particle resonance peak. The heavy renormalization and strong temperature dependence of the quasi-particle resonances can explain the large specific heat and gap-like resistivity found in Pu~\cite{PhysRevLett.91.205901}. Note that very recently Yee \emph{et al.} have performed DFT + DMFT calculations for plutonium chalcogenides and pnictides~\cite{PhysRevB.81.035105}. They declared that the combination of 5$f$ valence fluctuations and atomic multiplet structures might be responsible for the emergence of a multiplet of many-body quasi-particle peaks in Pu. Afterwards, they named these peaks as ``quasi-particle multiplets" and employed them to elucidate the observed photoemission triplet~\cite{joyce:2019,PhysRevB.62.1773}. On the other hand, A. B. Shick \emph{et al.} have proposed another similar concept ``Racah materials"~\cite{shick:2015} or ``Racah metals"~\cite{PhysRevB.87.020505}. In their scenario, the spectra of these materials usually contain two distinctive parts. Near the Fermi edge, there are well-pronounced atomic multiplet structures. However, for the rest part of the spectra, the multiplets are merged into single Hubbard band. They argued that $\delta$-Pu is a candidate of the so-called Racah metal, while PuB$_{6}$ and SmB$_6$ are supposed to be some kind of Racah materials (Racah insulators or Racah semiconductors). Just following their ideas, we find that the spectra of the $\alpha$, $\beta$, and $\gamma$ phases exhibit clear fingerprints of the quasi-particle multiplets (or Racah metals), while in the $\delta$, $\delta'$ and $\epsilon$ phases the quasi-particle multiplets dissolve due to high temperature.

\subsection{Valence state fluctuations and mixed-valence behaviors}

\begin{figure*}[ht]
\centering
\includegraphics[width=\textwidth]{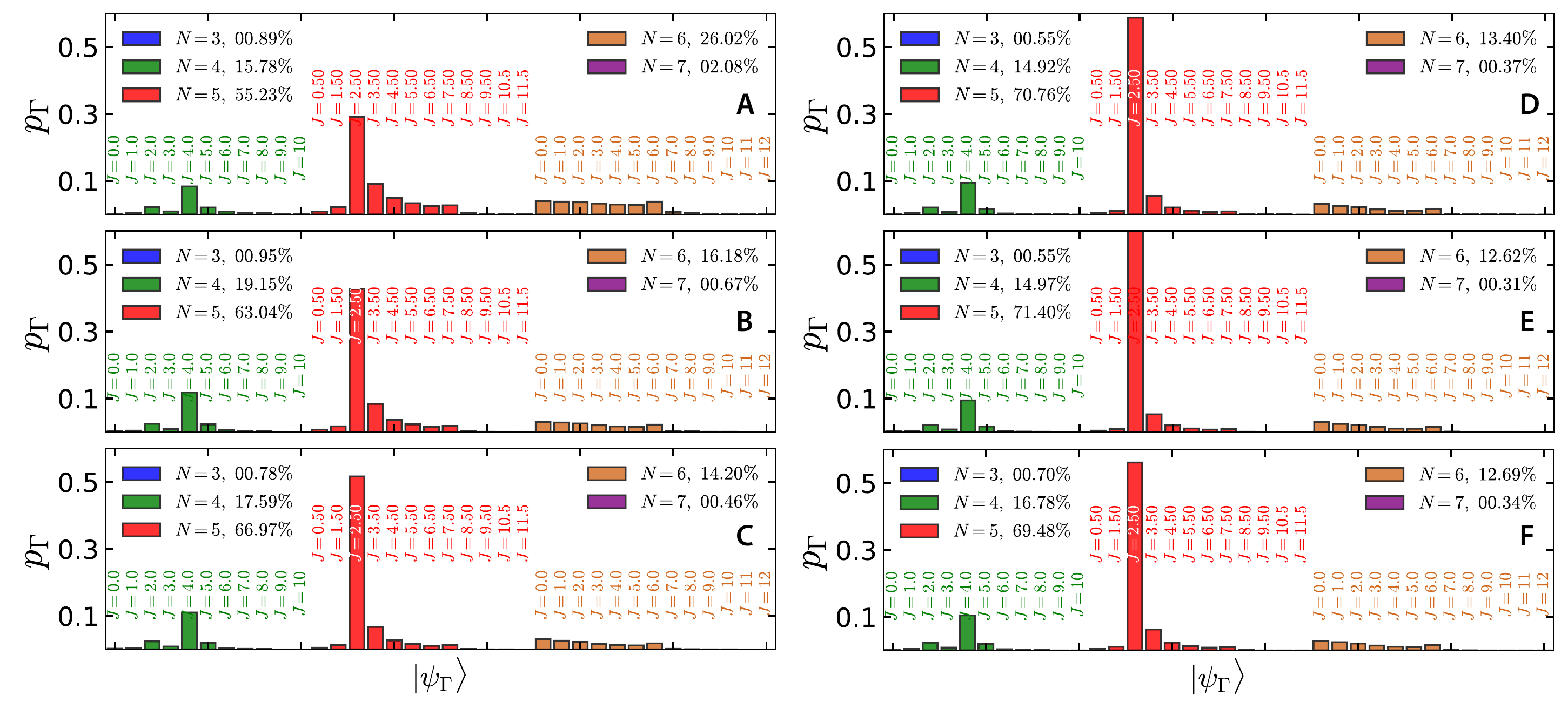}
\caption{(Color online). Valence state histograms of Pu by DFT + DMFT calculations. (a)-(f) Atomic eigenstate probabilities of $\alpha$, $\beta$, $\gamma$, $\delta$, $\delta'$, and $\epsilon$-Pu. The atomic eigenstates are denoted by using good quantum numbers $N$ (total occupancy) and $J$ (total angular momentum), i.e, $|\psi_\Gamma\rangle \equiv |N,J\rangle$. Note that data for the atomic eigenstates with ${N} = 3$ and ${N} = 7$ ($5f^3$ and $5f^7$ configurations) are not shown in these panels, because their contributions are too trivial ($< 1\%$) to be seen. The distributions of atomic eigenstate probabilities with respect to different $N$ are displayed in the legends. \label{fig:tprob}}
\end{figure*}

\begin{figure}[ht]
\centering
\includegraphics[width=\columnwidth]{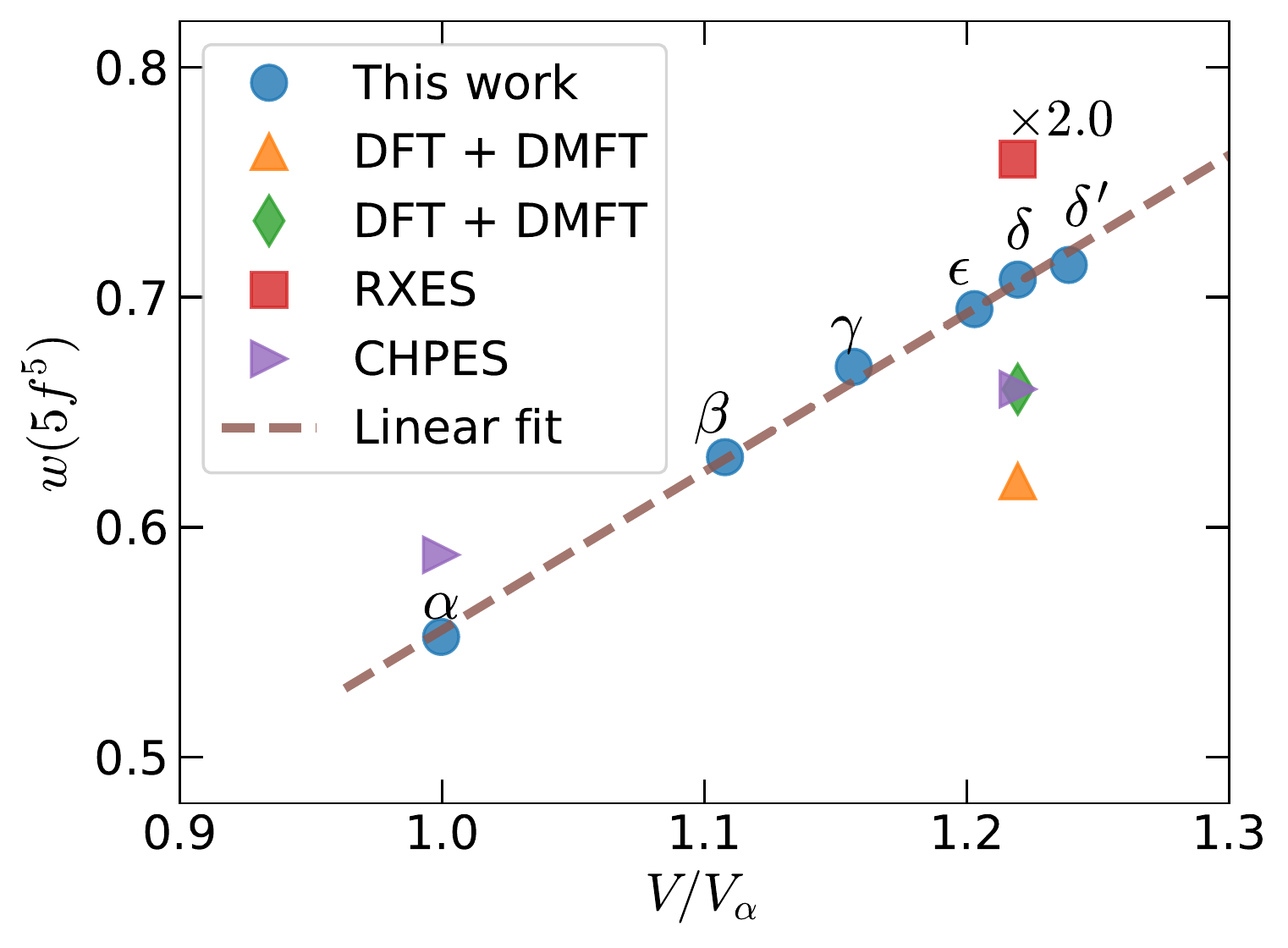}
\caption{(Color online). The $5f^{5}$ weight $w(5f^5)$ with respect to the unit cell volume $V$ of various phases of Pu. The old DFT + DMFT data for $\delta$-Pu are taken from Ref.~[\onlinecite{shim:2007}] (upper triangle symbols) and Ref.~[\onlinecite{Janoscheke:2015}] (diamond symbols). The data from resonant X-ray emission spectroscopy and core-hole photoemission spectroscopy are taken from Ref.~[\onlinecite{Booth26062012}] (square symbols) and Ref.~[\onlinecite{PhysRevB.82.045114}] (right triangle symbols), respectively. \label{fig:tvol}}
\end{figure}

Atomic eigenstate probability, or equivalently valence state histogram, has been already proven to be an useful observable to examine the valence state fluctuation or mixed-valence behavior in strongly correlated materials~\cite{PhysRevB.81.035105,shim:2007}. It represents the probability $p_{\Gamma}$ to find out a valence electron in given atomic eigenstates $|\psi_{\Gamma}\rangle$, which are usually labelled by using some good quantum numbers (such as $N$ or $J$)~\cite{PhysRevB.75.155113}. If valence electrons only favor one or two dominant atomic eigenstates (of course the corresponding atomic eigenstate probabilities are high), we can affirm that the valence state fluctuation in such a system is weak or restricted~\cite{PhysRevB.99.045109}. On the contrary, if valence electrons can live in a large number of atomic eigenstates (i.e., there are no predominant atomic eigenstates), the valence state fluctuation could be very strong~\cite{Lawrence_1981}.    

Plutonium is known to be a typical mixed-valence metal with $N_{5f} \sim 5.2$ which has been demonstrated theoretically~\cite{shim:2007,PhysRevX.5.011008} and experimentally~\cite{Booth26062012,PhysRevB.82.045114,Janoscheke:2015}. However, most of previous studies only focused on the $\alpha$ and $\delta$ phases because of their importance in military industry. We know almost nothing about the $5f$ valence state fluctuations for the other phases, and an unified picture for the mixed-valence behaviors of Pu is highly desired. Fortunately, $p_{\Gamma}$ is a direct output of the CT-HYB quantum impurity solver~\cite{shim:2007,PhysRevB.75.155113}. Thus, in the present work we are able to figure out the valence state fluctuations of Pu exhaustively for the first time. The calculated valence state histograms of plutonium are given in Fig.~\ref{fig:tprob}. It is noticed that the distributions of 5$f$ electronic configurations can be computed via the following equation,
\begin{equation}
w(5f^i) = \sum_N \sum_J \delta(N-i) p_{\Gamma}.
\end{equation} 
Here, $w(5f^i)$ denotes the weight of the $5f^{i}$ electronic configuration. And $i \in [3,7]$, because we only kept the contributions from those atomic eigenstates with $N \in [3,7]$ in the self-consistent calculations. The calculated values of $w(5f^i)$ are also shown in Fig.~\ref{fig:tprob} as legends. 

As a first glimpse, the 5$f$ valence state fluctuations are quite strong in all phases of plutonium. The $5f^4$, $5f^5$, and $5f^6$ electronic configurations have considerable contributions~\cite{PhysRevB.82.045114,shim:2007}. Among them, the $5f^5$ electronic configuration is of greatest importance. And in the $5f^{5}$ electronic configuration, the atomic eigenstate $|N = 5, J = 2.5\rangle$ is undoubtedly overwhelming. For the $5f^4$ and $5f^6$ electronic configurations, the principal atomic eigenstates are $|N = 4, J = 4.0\rangle$ and $|N = 6, J= 0.0\rangle$, respectively. Second, the $\alpha$ phase presents the strongest valence state fluctuation. In $\alpha$-Pu, the atomic state probability of $5f^{5}$ only accounts for 55.23\%, which is certainly smaller than the other phases. However, the contribution from the $5f^7$ electronic configuration is considerable ($\sim$ 2.08\%), which is much larger than the other phases ($< 0.7\%$). We believe that it is the small lattice volume per Pu atom and strong hybridization between the $5f$ and $spd$ electrons who are responsible for the enhancement of the 5$f$ valence state fluctuation in $\alpha$-Pu. Third, the valence state fluctuation in the $\delta'$ phase is the weakest. In $\delta'$-Pu, the contributions from the $5f^3$ and $5f^7$ electronic configurations are trivial (0.55\% and 0.31\%). And the proportion of its $5f^5$ electronic configuration is as high as 71.40\%, which is larger than the other phases. Overall, the strengths of valence state fluctuations in Pu are as follows: $\alpha$-Pu $>$ $\beta$-Pu $>$ $\gamma$-Pu $>$ $\epsilon$-Pu $>$ $\delta$-Pu $>$ $\delta'$-Pu. This trend is roughly reverse to the one of electronic correlation strengths $\mathcal{C}$~\cite{PhysRevB.87.045109,chantis:2009}, and is quite similar to the one of 5$f$ integrated spectral weights near the Fermi level of Pu [i.e., $I_{5f}$, see Eq.~(\ref{eq:i5f})]. Fourth, we attempted to plot the $w(5f^5)$ against the unit cell volume $V$ of Pu (see Fig.~\ref{fig:tvol}). Quite surprisingly, we find that $w(5f^5)-V$ exhibits a quasi-linear relation. The $w(5f^5)$ increases monotonically with respect to $V$. It indicates that the $w(5f^5)$ might be considered as a quantitative tool to measure the status of the $5f$ electrons of Pu~\cite{PhysRevB.99.045122}. Finally, the 5$f$ occupancy could be estimated via the following approximate relation,
\begin{equation}
\label{eq:n5f}
n_{5f} \approx 3w(5f^3) + 4w(5f^4) + 5w(5f^5) + 6w(5f^6) + 7w(5f^7).
\end{equation}
We would like to stress that since the distributions of electronic configurations $w(5f^i)$ are fairly different for the various phases of Pu, one would naturally expect that the averaged 5$f$ occupancies for these phases are dissimilar as well (see Table~\ref{tab:ratio}). We will discuss this issue in the following.  

\subsection{X-ray branching ratios and 5$f$ orbital occupancies}

\begin{table}[ht]
\caption{The X-ray absorption branching ratio $\mathcal{B}$ and 5$f$ occupancy $n_{5f}$ for the six allotropes of Pu. \label{tab:ratio}}
\begin{ruledtabular}
\begin{tabular}{rcccccc}
& \multicolumn{6}{c}{$\mathcal{B}$}\\
\cline{2-7}
Method & $\alpha$-Pu & $\beta$-Pu & $\gamma$-Pu & $\delta$-Pu & $\delta'$-Pu & $\epsilon$-Pu \\
\hline
DFT + DMFT\footnotemark[1]  & 0.752 & 0.774 & 0.774 & 0.780 & 0.778 & 0.779 \\
DFT + DMFT\footnotemark[2]  &       &       &       & 0.830 &       &       \\
DFT + DMFT\footnotemark[9]  &       & 0.795 &       & 0.795 &       &       \\
DFT + $G$\footnotemark[3]   & 0.844 & 0.859 & 0.860 & 0.891 &       & 0.862 \\
Experiments\footnotemark[4] & 0.842 &       &       & 0.847 &       &       \\
Experiments\footnotemark[5] & 0.813 &&&  && \\
\hline
& \multicolumn{6}{c}{$n_{5f}$\footnotemark[10]}\\
\cline{2-7}
Method & $\alpha$-Pu & $\beta$-Pu & $\gamma$-Pu & $\delta$-Pu & $\delta'$-Pu & $\epsilon$-Pu \\
\hline
DFT + DMFT\footnotemark[1]  & 5.37  & 5.25  & 5.27  & 5.24  & 5.18  & 5.19  \\
DFT + DMFT\footnotemark[2]  &       &       &       & 5.20  &       &       \\
DFT + DMFT\footnotemark[6]  &       &       &       & 5.04  &       &       \\
DFT + DMFT\footnotemark[9]  &       & 5.20  &       & 5.05  &       &       \\ 
DFT + $G$\footnotemark[3]   & 5.26  & 5.19  & 5.20  & 5.20  &       & 5.20  \\
Experiments\footnotemark[7] & 5.22  &       &       & 5.22  &       &       \\
Experiments\footnotemark[8] & 5.16  &       &       & 5.28  &       &       \\
\hline
& \multicolumn{6}{c}{$n_{5/2}$\footnotemark[10]}\\
\cline{2-7}
Method & $\alpha$-Pu & $\beta$-Pu & $\gamma$-Pu & $\delta$-Pu & $\delta'$-Pu &$\epsilon$-Pu \\
\hline
DFT + DMFT\footnotemark[1]  & 3.71  & 3.88  & 3.89  & 3.93  & 3.90  & 3.92  \\
DFT + DMFT\footnotemark[9]  &       & 4.07  &       & 4.03  &       &       \\
\hline
& \multicolumn{6}{c}{$n_{7/2}$\footnotemark[10]}\\
\cline{2-7}
Method & $\alpha$-Pu & $\beta$-Pu & $\gamma$-Pu & $\delta$-Pu & $\delta'$-Pu & $\epsilon$-Pu \\
\hline
DFT + DMFT\footnotemark[1]  & 1.67  & 1.37  & 1.38  & 1.31  & 1.28  & 1.28  \\
DFT + DMFT\footnotemark[9]  &       & 1.13  &       & 1.02  &       &       \\
\end{tabular}
\end{ruledtabular}
\footnotetext[1]{The present work. The 5$f$ impurity occupancy is calculated via the Matsubara Green's function $G(i\omega_n)$ [see Eq.~(\ref{eq:n5f_g})]. If we use the atomic state probability to evaluate the occupancy [see Eq.~(\ref{eq:n5f})], the results will be a little smaller.}
\footnotetext[2]{See Ref.~[\onlinecite{shim:2007}].}
\footnotetext[3]{See Ref.~[\onlinecite{PhysRevX.5.011008}]. $T = 0$\ K. The data for the $\alpha$ and $\beta$ phases are actually mean values for all of the non-equivalent atomic sites.}
\footnotetext[4]{See Ref.~[\onlinecite{PhysRevLett.93.097401}]. Using the electron energy-loss spectroscopy and X-ray absorption spectroscopy.}
\footnotetext[5]{See Ref.~[\onlinecite{PhysRevB.73.033109}]. Using the electron energy-loss spectroscopy.}
\footnotetext[6]{See Ref.~[\onlinecite{Janoscheke:2015}]. Using the atomic eigenstate probability.}
\footnotetext[7]{See Ref.~[\onlinecite{PhysRevB.82.045114}]. Using the core-hole photoemission spectroscopy.}
\footnotetext[8]{See Ref.~[\onlinecite{Booth26062012}]. Using the resonant X-ray emission spectroscopy.}
\footnotetext[9]{See Ref.~[\onlinecite{PhysRevB.99.125113}]. The data for the $\beta$ phase are actually mean values for all of the non-equivalent atomic sites.} 
\footnotetext[10]{$n_{5f} = n_{5/2} +n_{7/2}$.}
\end{table}

X-ray absorption spectroscopy is a powerful probe for the electronic transitions between core 4$d$ and valence 5$f$ states. The strong spin-orbital coupling for the 4$d$ states gives rise to two absorption lines, representing the $4d_{5/2} \rightarrow 5f$ and $4d_{3/2} \rightarrow 5f$ transitions, respectively~\cite{PhysRevLett.93.097401}. The X-ray absorption branching ratio $\mathcal{B}$ is defined as the relative strength of the $4d_{5/2}$ absorption line~\cite{PhysRevA.38.1943}. If the electrostatic interaction between core and valence electrons is skipped, the expression for $\mathcal{B}$ is given as follows~\cite{shim:2007,shim_epl09}:
\begin{equation}
\label{eq:ratio}
\mathcal{B} = \frac{3}{5} - \frac{4}{15} \frac{1}{14 - n_{5/2} - n_{7/2}} \left ( \frac{3}{2} n_{7/2} - 2 n_{5/2} \right ).
\end{equation}
Here, $n_{7/2}$ and $n_{5/2}$ are the 5$f$ occupation numbers for the $5f_{7/2}$ and $5f_{5/2}$ states, respectively. They can be calculated via the following equation:
\begin{equation}
\label{eq:n5f_g}
n_{\alpha} = \frac{1}{\beta} \sum_n e^{i\omega_n 0^{+}} G_{\alpha}(i\omega_n),
\end{equation} 
where $G_{\alpha}(i\omega_n)$ is the Matsubara Green's function and $\alpha$ is the orbital index. The X-ray absorption branching ratio $\mathcal{B}$ is a crucial physical quantity to represent the strength of the spin-orbital coupling interaction in the $f$ shell. It is usually extracted from the X-ray absorption spectroscopy and electron energy-loss spectroscopy or obtained via the atomic physics computations~\cite{RevModPhys.81.235,PhysRevLett.93.097401}. In order to gain a comprehensive insight into the interactions of 5$f$ electrons, in the present work we calculated $\mathcal{B}$ via the Eq.~(\ref{eq:ratio}) and Eq.~(\ref{eq:n5f_g}) additionally. The calculated results, together with the available experimental values~\cite{Janoscheke:2015,PhysRevB.82.045114,PhysRevLett.93.097401,PhysRevB.73.033109} and the DFT + $G$~\cite{PhysRevX.5.011008} (or DFT + DMFT~\cite{shim:2007,Janoscheke:2015,PhysRevB.99.125113}) results, are summarized in Table~\ref{tab:ratio}. 

Clearly, our DFT + DMFT values are marginally smaller than the experimental data, while the DFT + $G$ results overestimate $\mathcal{B}$. But, the accuracy of the experimental data is still questionable. For example, for $\alpha$-Pu, the two experimental values deviate from each other quite significantly~\cite{PhysRevLett.93.097401,PhysRevB.73.033109}. Nevertheless, we find that the expression $\mathcal{B}(\alpha) < \mathcal{B}(\beta) \sim \mathcal{B}(\gamma) < \mathcal{B}(\epsilon) \sim \mathcal{B}(\delta') \sim \mathcal{B}(\delta)$ approximately holds~\cite{PhysRevX.5.011008}. Similar (or reverse) trend has been identified in the electronic correlation strengths $\mathcal{C}$~\cite{chantis:2009} [or 5$f$ integrated spectral weights near the Fermi level $I_{5f}$, see Eq.~(\ref{eq:i5f})]. According to Eq.~(\ref{eq:ratio}), to calculate $\mathcal{B}$ not only the total 5$f$ occupancy $n_{5f}$, but also the orbital-resolved occupancies for the $5f_{5/2}$ and $5f_{7/2}$ states, i.e., $n_{5/2}$ and $n_{7/2}$, are essential inputs. These data are also collected and listed in Table~\ref{tab:ratio}. We find that the $\alpha$ phase has the largest 5$f$ occupancy ($n_{5f} \sim 5.37$). For the other phases, the $5f$ occupancy gets close to 5.2, which agrees quite well with the experiments~\cite{Booth26062012,PhysRevB.82.045114,PhysRevB.78.245101}. Note that $\alpha$ phase has the smallest $n_{5/2}$ ($\approx 3.71$) and the largest $n_{7/2}$ ($\approx 1.67$), while for the other phases $n_{5/2} \approx 3.90$ and $n_{7/2} \approx 1.30$. All these facts suggest that the electronic structure of the $\alpha$ phase is unique and different from the other phases. Besides, the $5f_{5/2}$ and $5f_{7/2}$ orbitals are six-fold and eight-fold degeneracies, respectively. So, the averaged occupancies per orbital are $\bar{n}_{5/2} \approx$ 0.62 and $\bar{n}_{7/2} \approx$ 0.21 for the $\alpha$ phase, and $\bar{n}_{5/2} \approx$ 0.65 and $\bar{n}_{7/2} \approx$ 0.16 for the other phases. This implies there exists nontrivial orbital differentiation in Pu's 5$f$ orbitals~\cite{PhysRevB.99.125113}. We will discuss this issue in the next subsection.

\subsection{Orbital-dependent 5$f$ electronic correlations}

\begin{figure}[ht]
\centering
\includegraphics[width=\columnwidth]{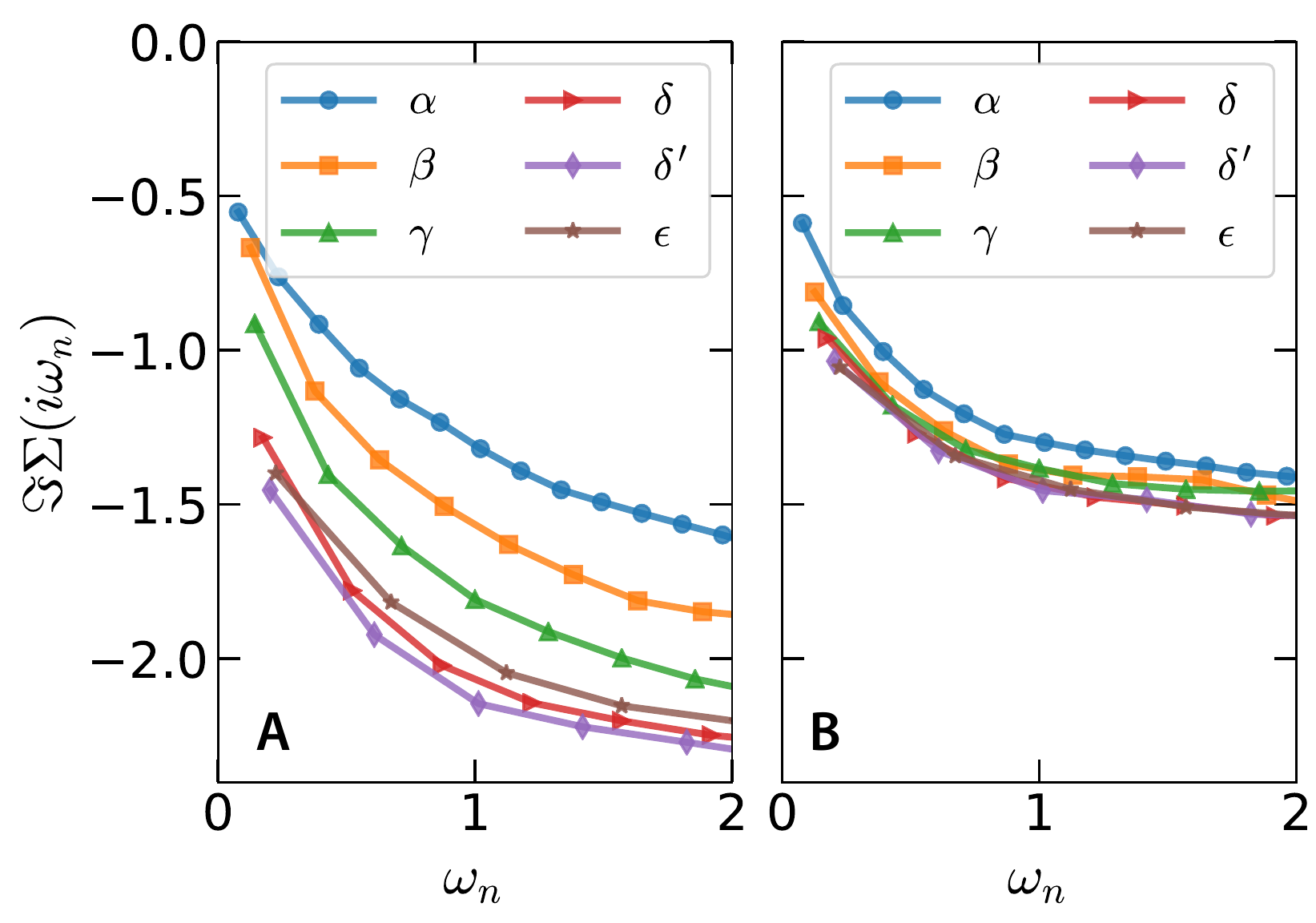}
\caption{(Color online). Imaginary parts of the Matsubara self-energy functions of Pu in the low-frequency regime by DFT + DMFT calculations. (a) $5f_{5/2}$ components. (b) $5f_{7/2}$ components. \label{fig:tsig_new}}
\end{figure}

\begin{table}[ht]
\caption{Calculated orbital-dependent quasiparticle weights $Z$ and electron effective masses $m^{*}$ for Pu. \label{tab:weight}}
\begin{ruledtabular}
\begin{tabular}{rcccccc}
 & $\alpha$ & $\beta$ & $\gamma$ & $\delta$ & $\delta$' & $\epsilon$ \\
\hline
$Z_{5/2}$\footnotemark[1]     & 0.125     & 0.158 & 0.135 & 0.120 & 0.122 & 0.138 \\
$Z_{7/2}$\footnotemark[1]     & 0.118     & 0.134 & 0.136 & 0.154 & 0.164 & 0.175 \\
$Z_{5/2}$\footnotemark[2]     &           & 0.07  &       & 0.05  &       & \\
$Z_{7/2}$\footnotemark[2]     &           & 0.21  &       & 0.28  &       & \\
$m^{*}_{5/2}$\footnotemark[1] & 8.03$m_e$ & 6.31$m_e$ & 7.39$m_e$ & 8.35$m_e$ & 8.17$m_e$ & 7.23$m_e$ \\
$m^{*}_{7/2}$\footnotemark[1] & 8.48$m_e$ & 7.46$m_e$ & 7.35$m_e$ & 6.51$m_e$ & 6.11$m_e$ & 5.70$m_e$ \\
$m^{*}_{5/2}$\footnotemark[2] &           & 14.0$m_e$ &           & 20.0$m_e$ &           &  \\
$m^{*}_{7/2}$\footnotemark[2] &           & 4.70$m_e$ &           & 3.57$m_e$ &           &  \\
\end{tabular}
\end{ruledtabular}
\footnotetext[1]{The present work. $Z$ and $m^*$ are evaluated using Eq.~(\ref{eq:weight}).}
\footnotetext[2]{See Ref.~[\onlinecite{PhysRevB.99.125113}]. Note that the one-crossing approximation (OCA) quantum impurity solver was used. So, $Z$ and $m^*$ were evaluated directly using the real-axis self-energy functions:
$Z^{-1} = 1 - \frac{\partial}{\partial \omega} \text{Re} \Sigma(\omega) \big|_{\omega = 0}.$
The data for the $\beta$ phase are actually averaged values for all of the non-equivalent atomic sites.} 
\end{table}

In general, all electronic correlations beyond the DFT level (single particle picture) can be encapsulated in self-energy functions. In Fig.~\ref{fig:tsig_new}, the Matsubara self-energy functions (only the imaginary parts at low frequency) for 5$f$ orbitals of Pu are shown. These self-energy functions show the following features. First of all, no doubt, the low-frequency parts of self-energy functions are concave, implying metallic solutions. Second, the intercept in $y$-axis means the low-energy electron scattering rate $\gamma$. We find that $\gamma$ is the smallest (largest) for the $\alpha$ ($\delta$') phase. Third, the $5f_{5/2}$ and $5f_{7/2}$ states exhibit quite different behaviors. The low-energy scattering rates of the $5f_{7/2}$ states are smaller than those of the $5f_{5/2}$ states. As for the $5f_{5/2}$ states, the self-energy functions for various Pu's allotropes are distinct. While for the $5f_{7/2}$ states, it is hardly to distinguish self-energy functions for the six phases of Pu. Finally, it seems that the self-energy functions at low frequency region is not linear, deviating from the prediction of Landau's Fermi-liquid theory~\cite{RevModPhys.68.13}. This might be a possible explanation for the bad metal behaviors of Pu observed below room temperature~\cite{handbook,HECKER2004429,LAReview}.            

We can further evaluate the quasiparticle weights $Z$ and electron effective masses $m^{*}$ through the following equation:
\begin{equation}
\label{eq:weight}
Z^{-1} = \frac{m^{*}}{m_e} \approx 1 - \frac{\text{Im} \Sigma(i\omega_0)}{\omega_0}.
\end{equation}
Here, $\omega_0 \equiv \pi / \beta$ and $m_e$ means the mass of the free band electron~\cite{RevModPhys.68.13}. We tried to calculate the orbital-dependent $Z$ and $m^{*}$ for the six allotropes of Pu. The results are collected in Table~\ref{tab:weight}. Though Eq.~(\ref{eq:weight}) might be not accurate enough at high temperature, we still have some interesting findings. First, the 5$f$ electrons in Pu are strongly correlated. The 5$f$ bands are strongly renormalized. The quasiparticle weights are between 0.1 and 0.2, and electron effective masses $m^{*}$ are between 5.0$m_e$ and 9.0$m_e$. Second, the $5f_{7/2}$ bands are more renormalized than the $5f_{5/2}$ bands at the high-temperature $\delta$, $\delta$' and $\epsilon$ phases, while they become less renormalized at the low-temperature $\alpha$, $\beta$, and $\gamma$ phases. Actually, we can define a new quantity, $R \equiv Z_{5/2} / Z_{7/2} $. We realize that $R > 1$ for $\alpha$- and $\beta$-Pu; $R \approx 1$ for $\gamma$-Pu; $R < 1$ for $\delta$-, $\delta$'-, and $\epsilon$-Pu. At last, the orbital differentiations (between the $5f_{5/2}$ and $5f_{7/2}$ states) are quite sizable, except for the $\gamma$ phase. It is suggested that $R$ is a good indicator to measure the orbital differentiation of 5$f$ orbitals. Lately, Brito \emph{et al.} have published DFT + DMFT results for $\beta$- and $\delta$-Pu~\cite{PhysRevB.99.125113}. Though they used a completely different quantum impurity solver based on one-crossing approximation~\cite{RevModPhys.78.865}, conspicuous orbital differentiations in $Z$ and $m^{*}$ were also observed. Thus, we can conclude that the 5$f$ electronic correlations in Pu are moderately orbital-dependent, and orbital-dependent 5$f$ electronic correlation may be a common characteristic in Pu~\cite{PhysRevB.99.125113} and the other actinide-based materials~\cite{PhysRevB.99.045109}. 


\section{Compared to the experimental results\label{sec:compare}}

\begin{figure}[ht]
\centering
\includegraphics[width=\columnwidth]{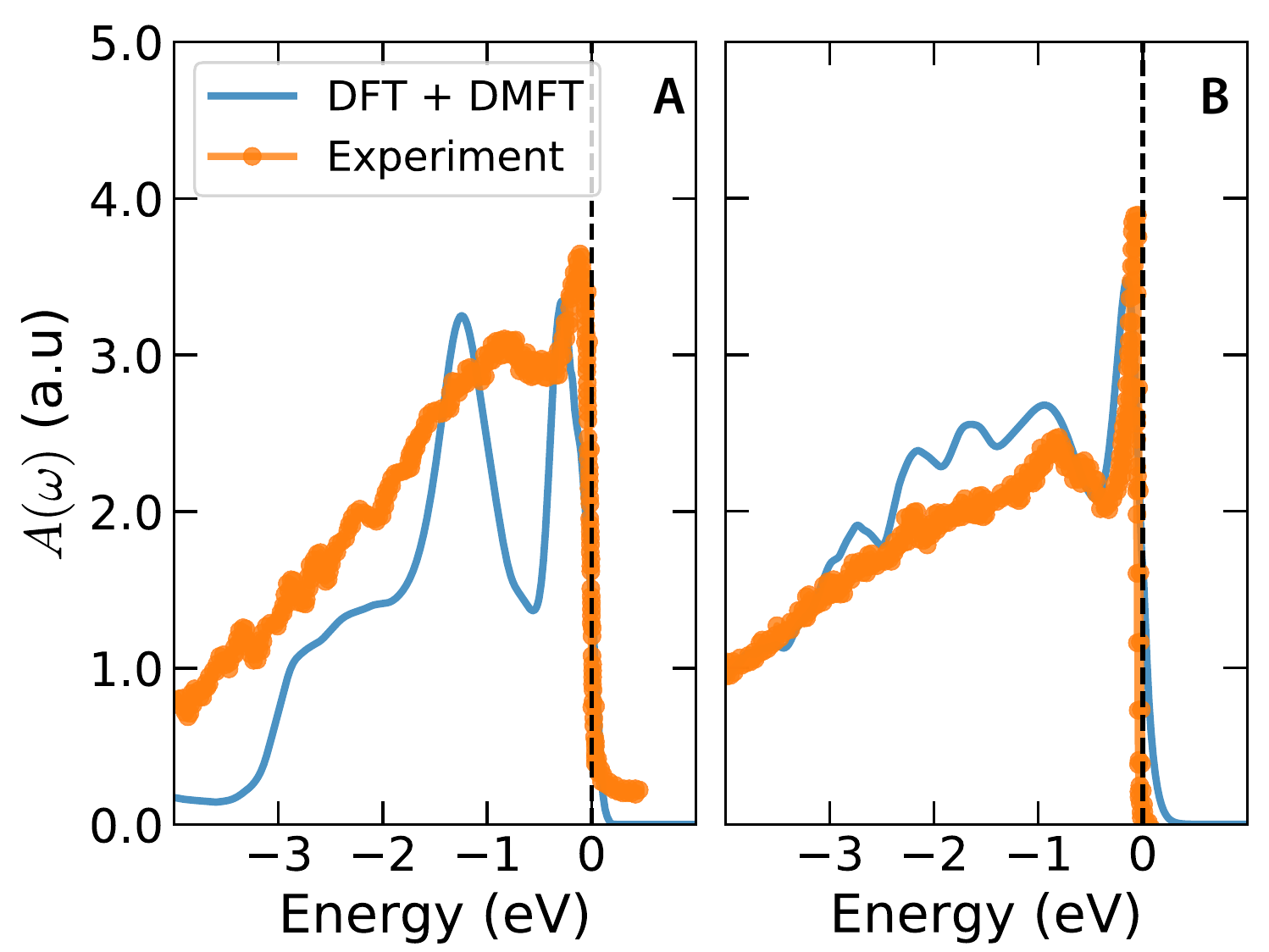}
\caption{(Color online). Comparisons of theoretical and experimental density of states for Pu. (a) $\alpha$-Pu. (b) $\delta$-Pu. The calculated spectra are represented as solid blue lines. The experimental data (filled orange circles) are taken from Ref.~[\onlinecite{PhysRevB.62.1773}]. The Fermi level $E_F$ is represented by vertical dashed line.  \label{fig:texp}}
\end{figure}

\begin{table}[ht]
\caption{Distributions of 5$f$ configurations for $\alpha$- and $\delta$-Pu. \label{tab:5flife}}
\begin{ruledtabular}
\begin{tabular}{rccccc}
& \multicolumn{5}{c}{$\alpha$-Pu} \\
\cline{2-6}
Method                     & $5f^{3}$ & $5f^{4}$ & $5f^{5}$ & $5f^{6}$ & $5f^{7}$ \\
\hline
DFT + DMFT\footnotemark[1] & 00.89\%  & 15.78\%  & 55.23\%  & 26.02\%  & 02.08\%  \\
Experiments\footnotemark[2]&          &   9.6\%  & 58.8\%   & 31.6\%   &          \\
Experiments\footnotemark[3]&          &  19.0\%  & 46.0\%   & 35.0\%   &          \\
\hline
& \multicolumn{5}{c}{$\delta$-Pu} \\
\cline{2-6}
Method                     & $5f^{3}$ & $5f^{4}$ & $5f^{5}$ & $5f^{6}$ & $5f^{7}$ \\
\hline
DFT + DMFT\footnotemark[1] & 00.55\%  & 14.92\%  & 70.76\%  & 13.40\%  & 00.37\%  \\
DFT + DMFT\footnotemark[4] &          & 7.5\%    & 62.1\%   & 30.4\%   &          \\
DFT + DMFT\footnotemark[5] &          & 12.0\%   & 66.0\%   & 21.0\%   &          \\
Experiments\footnotemark[2]&          &  5.7\%   & 66.4\%   & 27.8\%   &          \\
Experiments\footnotemark[3]&          & 17.0\%   & 38.0\%   & 45.0\%   &          \\
\end{tabular}
\end{ruledtabular}
\footnotetext[1]{The present work.}
\footnotetext[2]{See Ref.~[\onlinecite{PhysRevB.82.045114}]. Using the core-hole photoemission spectroscopy.}
\footnotetext[3]{See Ref.~[\onlinecite{Booth26062012}]. Using the resonant X-ray emission spectroscopy.}
\footnotetext[4]{See Ref.~[\onlinecite{shim:2007}].}
\footnotetext[5]{See Ref.~[\onlinecite{Janoscheke:2015}].}
\end{table}

It is well-known that plutonium is not only highly reactive, but also highly radioactive and toxic. Experimentally, working on plutonium metal demands special facilities. So, it is very difficult to conduct extensive experiments to study the electronic structures and the other physical properties of Pu~\cite{handbook,HECKER2004429,LAReview}. Basically, most of the calculated results presented above can be considered as critical predictions. In this section, we would like to compare our data with the experimental results (if available) and the previously theoretical results. We hope that these comparisons will enhance the rationality and significance of our predictions.

\emph{Band structures and spectral functions.} As far as we know, nowadays only the photoemission spectra for $\alpha$-Pu and $\delta$-Pu have been measured~\cite{PhysRevB.68.155109,PhysRevB.71.165101,joyce:2011,PhysRevB.62.1773,PhysRevB.65.235118,PhysRevB.75.035101,Tobin_2008,joyce:2019}. No ARPES experiments for Pu were reported in the public literatures. In Fig.~\ref{fig:texp}, we make a detailed comparison between the theoretical and experimental spectra. For $\alpha$-Pu, one has very good agreement between the experimental and theoretical spectra between -4.0 eV and the Fermi level. Especially, our spectrum shows a peak near -1.0 eV, which is consistent with the experiment~\cite{PhysRevB.62.1773} but contrary to the previous DFT + DMFT calculations which employed the simple $\mathcal{T}$-matrix fluctuation-exchange approximation as the quantum impurity solver~\cite{PhysRevB.75.235107}. For $\delta$-Pu, the calculated spectrum agrees quite well with the experiment in the vicinity of Fermi level. For lower energies below -0.5 eV, the agreement is less satisfactory. As already pointed out in both the DFT + DMFT studies~\cite{shim:2007} and photoemission experiments~\cite{PhysRevB.68.155109,PhysRevB.71.165101,joyce:2011,PhysRevB.62.1773,PhysRevB.65.235118,PhysRevB.75.035101,Tobin_2008,joyce:2019} of $\delta$-Pu, there are two additional satellite peaks below the Fermi level ($\omega \sim$ -0.5 and -1.0 eV). However, our spectral function only exhibits a weak single peak near -1.0 eV. Interestingly, we considered the atomic eigenstates with $N \in [3,7]$ in the present calculations. Nevertheless, if we try to restrict the atomic eigenstates to satisfy $N \in [4,6]$, the discrepancy between theory and experiment disappears and we can reproduce the double-peak structure near -1.0 eV (see Fig.~\ref{fig:t_dos_d}). Next we will discuss this issue in depth. Besides, we identify a few broad peaks around -2.0 eV in the spectral function. Though these peaks have been pointed out by Pourovskii \emph{et al.}~\cite{PhysRevB.75.235107}, Gorelov \emph{et al.}~\cite{PhysRevB.82.085117}, and Shim \emph{et al.}~\cite{shim:2007,PhysRevLett.101.126403} in their prior DFT + DMFT studies as well, they are all missing in the experimental spectra. According to Fig.~\ref{fig:tdos}, these peaks don't stem from the 5$f$ states. It is the $spd$ conduction states who make significant contributions to the photoemission spectra in this region. Finally, we would like to point out that the experimental spectra for $\alpha$-Pu and $\delta$-Pu are remarkably similar (though their crystal volumes, lattice structures, and mechanical properties are a bit different), besides $\delta$-Pu has a sharper and narrower Kondo peak than $\alpha$-Pu. However, we can easily distinguish the calculated spectral functions of $\alpha$-Pu and $\delta$-Pu, since the difference is very apparent. 

\emph{Valence state fluctuations.} Next, let us concentrate on the ground state 5$f$ weights of $\alpha$- and $\delta$-Pu. The calculated and experimental data for the proportions of $5f^{4}$-$5f^{7}$ configurations are listed in Table~\ref{tab:5flife}. For $\alpha$-Pu, overall the experimental data are in accordance with our prediction, besides the proportion of $5f^4$ is somewhat overestimated while those of $5f^{5}$ and $5f^{6}$ are slightly underestimated. For $\delta$-Pu, the available DFT + DMFT and experimental data are quite diverse. Some data deviate from the others apparently. For example, the data obtained via resonant X-ray emission spectroscopy suggest that the 5$f$ electronic configuration fractions for the $5f^4$, $5f^5$, and $5f^6$ states are 17\%, 38\%, and 45\%, respectively (see Fig.~\ref{fig:tvol} as well)~\cite{Booth26062012}. Though the $\delta$-Pu sample used in this experiment is not pure element (containing 1.9 at.\% Ga), such a large fraction for $5f^6$ configuration is hard to be understood and contrast to most of the DFT + DMFT~\cite{Janoscheke:2015,shim:2007} and experimental results~\cite{PhysRevB.82.045114}. Anyhow, our calculated results for $\delta$-Pu are excellently consistent with the very recent DFT + DMFT results~\cite{Janoscheke:2015}, and close to the experimental results obtained by using the core-hole photoemission spectroscopy~\cite{PhysRevB.82.045114}.  

\emph{Error analysis.} In general, our DFT + DMFT calculated and experimental results are reasonably consistent. But obvious deviations are present. We believe that these discrepancies between theory and experiment can be explained by the following reasons: (i) Temperature effect. The photoemission experiments for $\alpha$-Pu and $\delta$-Pu have been done at the same temperature 80 K~\cite{PhysRevB.62.1773}, while we carried out our DFT + DMFT calculations for $\alpha$-Pu and $\delta$-Pu at 290 K and 645 K (see Table~\ref{tab:param}), respectively. (ii) Mixture of $\alpha$ and $\delta$ phases. As a matter of fact, pure $\delta$-Pu is metastable or even unstable below room temperature and the Pu's $\delta-\alpha$ transition easily occurs~\cite{HECKER2004429}. In other words, one sample is supposed to be pure $\delta$ phase, but it is actually a mixture of $\alpha$ and $\delta$ phases. As mentioned before, Ga can stabilize the $\delta$ phase at low temperature. However, due to the segregation effect of Ga atom, the surface of $\delta$ phase Pu-Ga alloy usually tends to form $\alpha$ phase like structure~\cite{LAReview}. (iii) Limitations of DFT + DMFT calculations. In order to let the computational resources be affordable, we make some approximations in the calculations. A few approximations are severe. For example, we ignore the inequivalent Pu atoms in the $\alpha$-, $\beta$-, and $\gamma$-Pu, and truncate the atomic eigenstates (only those with $N \in [3,7]$ are kept). Even we used the same Coulomb interaction parameters ($U$ and $J_{\text{H}}$) for various phases of Pu (see Table~\ref{tab:param} for more details). These assumptions and approximations are probably major sources of error and uncertainty of our calculated results.


\section{Discussion\label{sec:discuss}}

In this section, we would like to address several important issues and problems.

\subsection{Evolution of 5$f$ electron localization}

According to the calculated results, we find that the 5$f$ electronic structures of the six allotropes of Pu share a lot of features. For example, the 5$f$ electrons are strongly correlated with large electron effective masses $m^*$ and small renormalization factors $Z$. They are in the midway of completely itinerant and localized. The electronic correlation is orbital-dependent. However, this is not the full story. The 5$f$ electronic structures of various phases of Pu are quite different. On one hand, the low-temperature and low-symmetry phases ($\alpha$, $\beta$, and $\gamma$ phases) are likely typical Racah metals. They exhibit quasiparticle multiplets in the density of states at the Fermi level. Their 5$f$ electrons favor the itinerant state more or less. The valence state fluctuation and mixed-valence behavior are quite remarkable, especially in the $\alpha$ phase. On the other hand, in the high-temperature and high-symmetry phases ($\delta$, $\delta$', and $\epsilon$ phases), the quasiparticle multiplets are merged into a single Kondo resonance peak. The 5$f$ electrons become more localized and the valence state fluctuation are somewhat restrained, especially in the $\delta$ and $\delta$' phases. In $\delta$'-Pu, the proportion of 5$f^5$ configuration is the largest, while the percentages of 5$f^4$ and 5$f^{6}$ configurations are the smallest. These results are consistent with such a fact that the $\delta'$ phase has the largest atomic volume and the smallest density when extrapolated to zero temperature~\cite{HECKER2004429} (and thus its 5$f$ electrons are more close to fully localized)~\cite{PhysRevX.5.011008}. From the $\alpha$ to $\delta$ ($\delta$') phases, the lifetime for $5f^{5}$ states increases while the ones for $5f^{4}$ and $5f^6$ states decrease, we also expect a crossover for 5$f$ electrons from itinerant state to partially localized state.     

\subsection{Truncation approximation for atomic eigenstates}

\begin{figure}[ht!]
\centering
\includegraphics[width=\columnwidth]{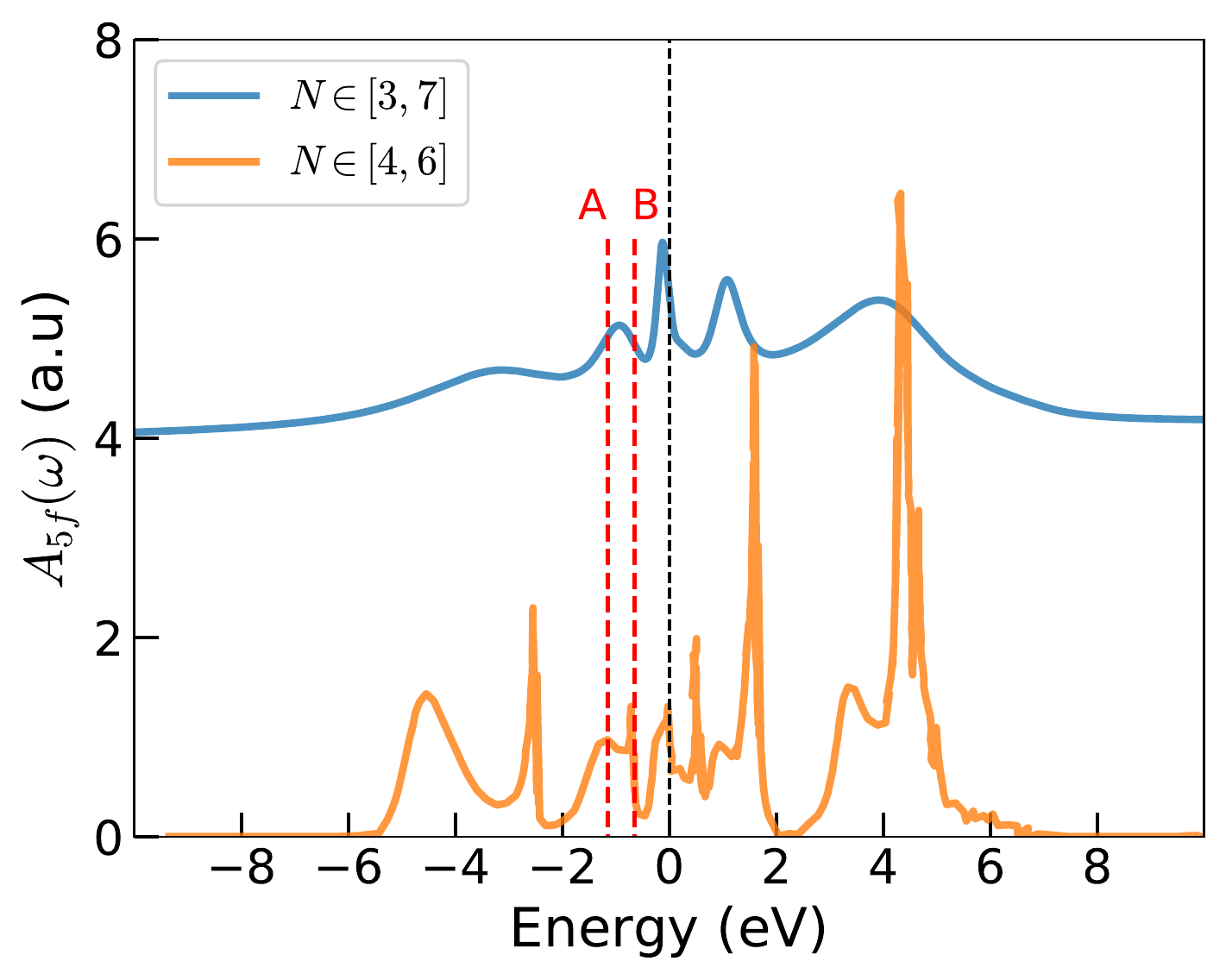}
\caption{(Color online). 5$f$ density of states for $\delta$-Pu. Here, the spectra obtained by considering the $N \in [3,7]$ and $N \in [4,6]$ atomic eigenstates are represented as blue and orange solid lines, respectively. The former is shifted upward as a whole. The vertical dashed line (black) means the Fermi levels $E_F$. In this figure, the two-peak structure between -2 and 0 eV is labelled (A and B). \label{fig:t_dos_d}}
\end{figure}

In the present DFT + DMFT calculations, we employed the numerically exact CT-HYB quantum Monte Carlo algorithm as quantum impurity solver to solve the Anderson impurity models for 5$f$ electrons. The most time-consuming part for the CT-HYB quantum impurity solver is to compute the local trace which involves a sequence of matrix multiplications between the time evolution operators and creation/annihilation operators for the impurity electrons~\cite{PhysRevB.75.155113,PhysRevLett.97.076405,RevModPhys.83.349}. For the $5f$ shell, the size of operator matrix is $2^{14} \times 2^{14}$ which requires huge computer memory to save the data, and one typically needs to multiply a few hundred of these matrices at each Monte Carlo step. We couldn't imagine how to simulate this without any approximations. In order to overcome this bottleneck, we chose some good quantum numbers such as $N$ and $J$ to divide the whole Hilbert space into sub-blocks to meet the memory limit~\cite{PhysRevB.75.155113}, and then utilized the Lazy trace evaluation trick~\cite{PhysRevB.90.075149} to accelerate the calculation. Even though these strategies are used, the calculations are still unaffordable. Certainly, we need to adopt more aggressive approximations. In the present work, we considered the truncation approximation for the atomic eigenstates to gain further acceleration. The truncation we adopted is in relation to the nominal occupancy $N$ of atomic eigenstates. Explicitly, only those atomic eigenstates whose occupancy $N$ satisfy $N \in [N_{\text{low}}, N_{\text{high}}]$ will be taken into accounts in the local trace evaluation. Obviously, though such a truncation will improve the computational efficiency greatly, it will introduce some uncontrollable biases at the same time. Therefore, we have to evaluate carefully how large the discrepancies are due to this severe truncation.

As for Pu, the situation is in a dilemma. The nominal 5$f$ occupancy is about 5, so one of the most radical truncations is to consider the $N \in [4,6]$ atomic eigenstates, which can save a lot of computer resources indeed. A somewhat safe choice is to retain the $N \in [3,7]$ atomic eigenstates, but it consumes much more memories and CPU hours. Which one is better? At first, let's go back to Fig.~\ref{fig:tprob}. We discover that the contributions from the $N = 3$ and $N = 7$ atomic eigenstates (i.e., the $5f^3$ and $5f^7$ electronic configurations) are considerable. Specially, for $\alpha$-Pu and $\beta$-Pu, the contributions from the $N = 7$ atomic eigenstates are 2.08\% and 0.67\%, respectively, which couldn't be simply ignored. In $\alpha$-Pu and $\beta$-Pu, the 5$f$ electrons are less localized and the valence state fluctuations are more conspicuous. Hence the $N = 3$ and $N = 7$ atomic eigenstates are more important for them. Second, let's focus on the density of states of $\delta$-Pu again. We recalculated it with two different truncations, i.e., $N \in [4,6]$ and $N \in [3,7]$. The results are compared in Fig.~\ref{fig:t_dos_d}. The spectra obtained within $N \in [4,6]$ truncation show multiple sharp peaks near $E_F$ which are associated with the atomic multiplets. Particularly, in the energy range of [-2~eV,~0~eV], we observed two additional peaks besides the Kondo resonance peak at $E_{F}$. The two peaks belong to the $5f_{5/2}$ and $5f_{7/2}$ states, respectively, and are marked with vertical red lines in Fig.~\ref{fig:t_dos_d}. Note that the two peaks were already identified in the early DFT + DMFT calculations~\cite{shim:2007} and photoemission experiments~\cite{PhysRevB.62.1773,joyce:2011,PhysRevB.71.165101,PhysRevB.68.155109,Tobin_2008,PhysRevB.65.235118,PhysRevB.75.035101}. On the contrary, if we consider more atomic eigenstates with $N \in [3,7]$, the calculated spectrum looks slightly different. In the spectrum, the peaks from the atomic multiplets are replaced with a broad ``hump". The Kondo resonance peak still exists, but the two additional peaks between -2.0 eV and 0.0 eV are merged into a single shoulder peak which is consistent with the very recent DFT + DMFT results~\cite{Janoscheke:2015}. Since the more atomic eigenstates are included in the calculations, the more virtual charge fluctuations between different atomic eigenstates contribute to the spectrum. Finally, the new spectrum will become broader, and more and more featureless. Actually, it just looks like an envelop of the old one on the whole. Nevertheless, we believe that even for $\delta$- and $\delta$'-Pu which have more localized 5$f$ electrons than $\alpha$- and $\beta$-Pu, the influences from the $N = 3$ and $N = 7$ atomic eigenstates are still remarkable. So it is essential to keep them in the DFT + DMFT calculations for Pu at a cost of greatly increasing computational resource consumptions.

\subsection{Negative sign problem}

\begin{figure}[ht]
\centering
\includegraphics[width=\columnwidth]{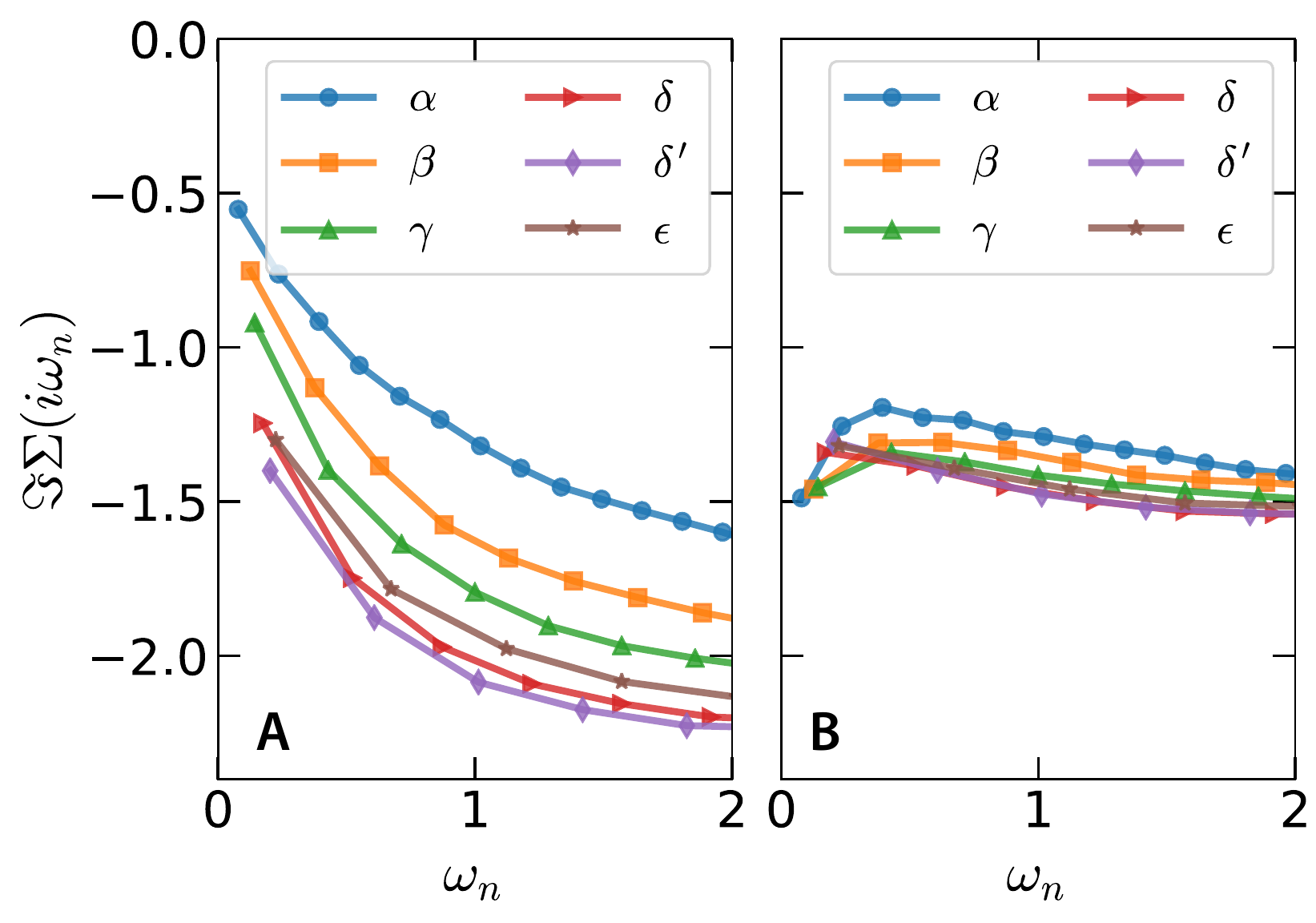}
\caption{(Color online). Imaginary parts of the Matsubara self-energy functions of Pu in the low-frequency regime by DFT + DMFT calculations. (a) $5f_{5/2}$ components. (b) $5f_{7/2}$ components. Note that in these DFT + DMFT (CT-HYB) calculations, the weights of Monte Carlo sampling are fixed to be positive. \label{fig:tsig}}
\end{figure}

In the present work, we chose CT-HYB as quantum impurity solver, which is based on a stochastic sampling of a perturbation expansion in the impurity-bath hybridization parameter~\cite{PhysRevB.75.155113}. Note that the CT-HYB quantum impurity solver is not sign-problem-free. In other words, for a given term in the perturbation expansion series, the weight might be not always positive. This is the well-known negative sign problem for fermionic quantum Monte Carlo algorithms~\cite{RevModPhys.83.349}. It will deteriorate the computational accuracy tremendously. The negative sign problem even gets worse when spin-orbital coupling is activated, system temperature and crystal symmetry are lowered, and off-diagonal terms in hybridization function emerge. In our calculations, we find that the averaged signs $\langle s\rangle$ are around 0.50 ($\alpha$), 0.52 ($\beta$), 0.61 ($\gamma$), 0.76 ($\delta$), 0.82 ($\delta$'), 0.82 ($\epsilon$). These are still acceptable. But, for the $\alpha$, $\beta$, and $\gamma$ phases, the negative sign problems are almost out of control. 

We have to increase the computational time as much as possible to relieve the impact of negative sign problem. An alternative method is to ignore the negative weights completely. That is to say, we can enforce the weights of Monte Carlo sampling to be positive. In order to validate this idea, we performed benchmark calculations with the same parameters and setups (see Table~\ref{tab:param}). Most of the calculated results, such as band structures, density of states, and valence state histograms are similar. It seems that they are not sensitive to the negative sign problem. However, we find some extraordinary phenomena in the Matsubara self-energy functions. In Fig.~\ref{fig:tsig}, new Matsubara self-energy functions are shown. They were obtained in the benchmark calculations without negative sign problem. Compared to the Matsubara self-energy functions shown in Fig.~\ref{fig:tsig_new}, $\text{Im} \Sigma_{5/2}(i\omega_n)$ is quite close. But the low-frequency behaviors of $\text{Im} \Sigma_{7/2}(i\omega_n)$ for the $\alpha$, $\beta$, and $\gamma$ phase are a bit exceptional. Their $\text{Im} \Sigma_{7/2}(i\omega_n)$ increase at first, reach maximum values when $\omega_n \approx 0.4$, and then decrease gradually with respect to $\omega_n$. Note that the negative sign problem is much more severe in these phases than the high-temperature phases. Clearly, ignoring the negative sign problem in the present cases leads to unphysical Matsubara self-energy functions. And in a very recent DFT + DMFT study for the $\beta$ phase, similar Matsubara self-energy functions like those shown in Fig.~\ref{fig:tsig} are observed as well. So, we believe that the negative sign problem must be carefully taken into consideration once the CT-HYB quantum impurity solver is adopted in the DFT + DMFT studies of $f$-electron systems (lanthanides and actinides), in which the spin-orbital coupling is strong and the crystal structures are usually quite complex. 

\subsection{Site-dependent electronic structures}

As mentioned before, the $\alpha$, $\beta$, and $\gamma$ phases contain multiple non-equivalent Pu atoms. In principle, each non-equivalent Pu atom is described by a unique quantum impurity model, which should be solved individually in the framework of DFT + DMFT approach~\cite{RevModPhys.78.865,PhysRevB.81.195107}. Despite that the CT-HYB quantum impurity solver is already the most powerful established impurity solver so far, solving the quantum impurity problems for 5$f$ electrons (Pu) is still extremely memory-consuming and time-consuming owing to the exponentially increasing Hilbert space and severe negative sign problem. So to study these low-symmetry phases using the DFT + DMFT approach without any simplifications becomes an impossible task. This is also the major reason why most of the previous DFT + DMFT calculations concerning with Pu metal were conducted for the high-symmetry $\delta$ phase. Actually, only a few years ago, Zhu \emph{et al.}~\cite{zhu:2013} reported the first DFT + DMFT calculations for $\alpha$-Pu. They have discovered the site-resolved electronic structures. To the best of our knowledge, it is the first time and perhaps the unique one to address the electronic structures represented by non-equivalent atoms in $\alpha$-Pu by employing charge fully self-consistent DFT + DMFT calculations. Their calculations cost huge computational resources (288 CPU cores, 1152 GB memory, and $>$ 2000 non-interrupted CPU wall-clock hours). The latest advances were made by Brito \emph{et al.}, who studied the site-dependent electronic structures in the $\beta$ phase~\cite{PhysRevB.99.125113}. They employed the OCA quantum impurity solver, which is faster than CT-HYB, but is not numerically exact~\cite{RevModPhys.78.865}. To carry out similar calculations for all allotropes of Pu is far beyond the ability of computational resources we owned. For this reason, in the present work, we have to restrict ourselves to consider only the completely degenerated Pu atoms. This assumption simplifies the calculations greatly, but undoubtedly leads to deviations to some extent. Zhu and Brito \emph{et al.}~\cite{zhu:2013,PhysRevB.99.125113} have revealed weak site dependence in the electronic structures of $\alpha$-Pu and $\beta$-Pu. We thus expect that in the $\gamma$ phase, the 5$f$ electronic structure would exhibit some kinds of non-trivial site-dependent features. This is still an open and interesting question. We will reexamine it in the future.

\subsection{Similarities and differences between the $f$ electronic structures of Ce and Pu}

We already knew, the mechanical and lattice dynamical properties of Ce and Pu are somewhat similar~\cite{alex:2012,koskenmaki1978337,PhysRevLett.92.146401,wong:2003,PhysRevB.79.052301,dai:2003,krisch:2011,huang:2007,PhysRevB.19.5746,PhysRevB.25.6485,PhysRevLett.108.045502}. The electronic structures of Ce and Pu share many similarities as well~\cite{PhysRevB.99.045122,Lu_2018}. Here we will attempt to summarize them as follows. First, both Ce and Pu are mixed-valence strongly correlated metals with non-integer 4$f$ and 5$f$ occupations. Second, the low-temperature phases ($\alpha$-Ce, $\beta$-Ce, $\alpha$-Pu, $\beta$-Pu, and $\gamma$-Pu) show stronger valence state fluctuations and weaker $f$ electronic correlation strengths. On the contrary, in the high-temperature phases ($\gamma$-Ce, $\delta$-Ce, $\delta$-Pu, $\delta'$-Pu, and $\epsilon$-Pu), $f$ electrons tend to be more localized and manifest stronger electronic correlations. From low temperature to high temperature, the $f$ electrons become more and more incoherent, and a itinerant-localized crossover might emerge. Third, the $4f$ ($5f$) electronic correlations are orbital-dependent. Finally, the $4f^0$, $4f^1$, and $4f^2$ final states in Ce correspond to the $5f^4$, $5f^5$, and $5f^6$ final states in Pu~\cite{PhysRevB.71.165101}. The photoemission spectra of Ce and Pu display similar features, i.e., the quasiparticle resonance peaks ($4f^1 \to 4f^2$, $5f^5 \to 5f^6$) at low binding energy and broad Hubbard bands ($4f^0 \to 4f^1$, $5f^4 \to 5f^5$) at high binding energy.

The most remarkable difference for the electronic structures of Ce and Pu is that there are quasiparticle multiplets in the low-temperature phases of Pu, while they are absent in Ce~\cite{PhysRevB.99.045122}. The quasiparticle multiplets will strongly affect the optical conductivity, resistivity, specific heat, and many other physical properties of Pu~\cite{PhysRevLett.91.205901}. Besides, the low-temperature and low-symmetry phases of Pu are supposed to exhibit nontrivial site-dependent electronic structures~\cite{zhu:2013,PhysRevB.99.125113}. However, for Ce, this possibility is already excluded theoretically. Very recently, we examined the site dependence of the 4$f$ electronic structure in the $\beta$ phase of Ce, which has two non-equivalent Ce atoms, via DFT + DMFT calculations. The calculated results suggest that it does not exhibit a site-selective 4$f$ localized state~\cite{PhysRevB.99.045122,Lu_2018}, opposite to our assumption.


\section{Concluding remarks\label{sec:summary}}

In the present paper, we employed the \emph{ab initio} many-body approach, namely the charge fully self-consistent DFT + DMFT method, to investigate the 5$f$ electronic structure of strongly correlated Pu metal. We endeavored to calculate the momentum-resolved spectral functions, total and $5f$ partial density of states, histograms of atomic eigenstates, X-ray branching ratios, 5$f$ orbital occupancies, and Matsubara self-energy functions for the six allotropes of Pu under ambient pressure. On one hand, the calculated results are in well consistent with the available experimental results. On the other hand, most of the calculated results presented in this paper can be regarded as essential predictions and require further experimental or theoretical examinations. The major findings of this work are as follows: (1) $\alpha$-, $\beta$-, and $\gamma$-Pu belong to the so-called Racah metals~\cite{PhysRevB.87.020505,shick:2015,PhysRevB.99.125113} which show quasiparticle multiplets~\cite{PhysRevB.81.035105} near the Fermi level in the spectral functions. While in the high-temperature phases ($\delta$-, $\delta$'-, and $\epsilon$-Pu), the quasiparticle multiplets merge into a single Kondo resonance peak. (2) Plutonium is a typical mixed-valence metal. Its valence state fluctuation is the strongest in the $\alpha$ phase, and the weakest in the $\delta'$ phase. (3) The 5$f$ electronic correlation is orbital dependent. We define a new variable $R$ to account for the 5$f$ orbital differentiation. Further analysis reveals that the $5f_{5/2}$ bands are more renormalized in the $\delta$, $\delta'$, and $\epsilon$ phases. While in the $\alpha$ and $\beta$ phases, so do the $5f_{7/2}$ bands. (4) The 5$f$ electrons in $\delta$'-Pu is the most localized, which matches up the fact that $\delta$'-Pu has the largest atomic volume when extrapolated to zero temperature~\cite{PhysRevX.5.011008}. (5) In order to obtain reliable results, we must retain the contributions from $N = 3$ and 7 atomic eigenstates, and consider the negative sign problem explicitly. (6) The site dependence of 5$f$ electronic structures for the $\alpha$, $\beta$, and $\gamma$ phases is probably nontrivial. Finally, we highlight the differences and similarities between Ce-4$f$ and Pu-5$f$ electronic structures. These calculated results support the conjecture that Pu lies on a knife edge of 5$f$ electron localization, and the six allotropes of Pu are totally different metals~\cite{PhysRevB.84.064105}.

We have to admit that there are some limitations and simplifications in our calculations. For instance, we ignore the site dependence of $5f$ electronic structures in the low-temperature phases of Pu, we also make severe truncations in treating the contributions of atomic eigenstates, the crystal structures are not optimized, the phase transition and phase stability of the six allotropes of Pu are not discussed, and so on. We will try to overcome these challenges and problems in the future. Nevertheless, in the present calculations, we not only reproduce the experimental results and provide some useful supplements to the experiments, but also discover some new physics and enrich our understanding about the extraordinary properties of Pu. Our work demonstrates again that the state-of-the-art DFT + DMFT method can be applied to study the intricate and delicate 5$f$ electronic structures of strongly correlated actinide metals quantitatively, shedding new light on the \emph{ab initio} calculations for lanthanides and actinides. Besides Pu, notice that the electronic structures for most of the other actinide metals (such as Pa, U, Np, Am, Cm, Bk, and Cf) remain unclear. These elements show complex phase diagrams and phase transitions as a function of temperature and pressure~\cite{RevModPhys.81.235,Heathman110,PhysRevB.63.214101,PhysRevB.87.214111,PhysRevLett.85.2961}. Therefore, it would be highly desired to apply the DFT + DMFT method to survey their lattice properties and electronic structures in the near future.  


\begin{acknowledgments}
We thank Dr. Yilin Wang for fruitful discussion. This work was supported by the Natural Science Foundation of China (No.~11504340 and No.~11704347), the Foundation of President of China Academy of Engineering Physics (No.~YZ2015012), and the Science Challenge Project of China (No.~TZ2016004).
\end{acknowledgments}


\bibliography{pu}

\begin{thebibliography}{122}
\bibitem [{lim()}]{limits}%
  \BibitemOpen
  \href@noop {} {}\bibinfo {note} {The upper and lower limits for the integral
  in Eq.~(\ref{eq:w5f}) are a bit arbitary. Here, they are determined by the
  width of the quasi-particle resonance peak at half height in the $\alpha$
  phase. That is the reason why we set them to 0.2 and -0.2.}\BibitemShut
  {Stop}%
\end{thebibliography}

\end{document}